\documentclass[journal=jchemmat]{achemso}

\usepackage[version=3]{mhchem} 

\usepackage{setspace}
\usepackage{xcolor}
\usepackage{times}
\usepackage{amsmath}
\usepackage{mathtools}
\usepackage{amssymb}
\usepackage[T1]{fontenc}
\usepackage{url}
\usepackage{bm}
\usepackage{here}
\usepackage{graphicx}
\usepackage{framed}
\definecolor{shadecolor}{rgb}{0.9,0.9,0.9}
\usepackage[utf8x]{inputenc}
\usepackage{textgreek}
\usepackage{tabularx}

\usepackage[below]{placeins}

\newlength{\figurewidth}
\author{Stefano Racioppi}
\affiliation[Buffalo]
{Department of Chemistry, State University of New York at Buffalo, Buffalo, NY, USA}
\author{Maosheng Miao}
\affiliation[Northridge]
{Department of Chemistry and Biochemistry, California State University Northridge, Northridge, CA, United States}
\author{Eva Zurek}
\affiliation[Buffalo]
{Department of Chemistry, State University of New York at Buffalo, Buffalo, NY, USA}
\email{ezurek@buffalo.edu}

\title[An \textsf{achemso} demo]
  {Intercalating Helium into A-site Vacant Perovskites}

\begin{document}

\begin{tocentry}
\begin{figure}[H] 
\includegraphics[width=8.4cm]{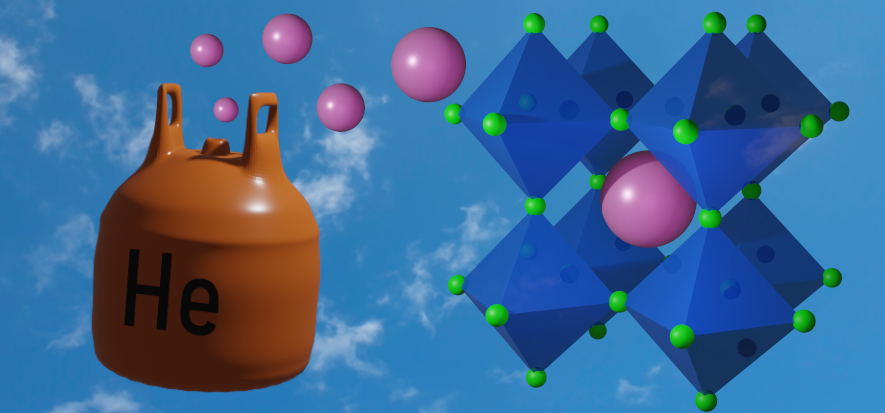}
			\label{fig:TOC}
\end{figure}





\end{tocentry}

\begin{abstract}
Evolutionary searches were employed to predict the most stable structures of perovskites with helium atoms on their A-sites up to pressures of 10~GPa. 
The thermodynamics associated with helium intercalation into [CaZr]F$_6$, structure that [He]$_2$[CaZr]F$_6$ adopts under pressure, and the mechanical properties of the parent perovskite and helium-bearing phase were studied via density functional theory (DFT) calculations. The pressure-temperature conditions where the formation of HeAlF$_3$, HeGaF$_3$, HeInF$_3$, HeScF$_3$ and HeReO$_3$ from elemental helium and the vacant A-site perovskites is favored were found. Our DFT calculations show that entropy can stabilize the helium-filled perovskites provided that the volume that the noble gas atom occupies within their pores is larger than within the elemental solid at that pressure. We find that helium incorporation will increase the bulk modulus of AlF$_3$ from a value characteristic of tin to one characteristic of steel, and hinders rotations of its octahedra that occur under pressure.
\end{abstract}

\newpage

\section{Introduction}
Helium, whose abundance in the Universe is only rivaled by hydrogen, came into existence during the Big Bang nucleosynthesis that occurred moments after the Cosmos began. Billions of years later the extreme pressures and temperatures found in the stars continue to be a source of helium, which is produced within them via fusion reactions. As a result, this second lightest element is prevalent in the stars, in gas-giants, and it may be stored deep within our planet.\cite{Jackson2010_Nature, Hubbard_2016_TheAstrophysJourn} However, helium is relatively rare in the Earth's atmosphere and its shortage impacts the deployment of weather balloons, cooling of medical and scientific equipment, and children's birthday parties. Part of the reason for this terrestrial shortage is helium's inertness -- it is the least reactive of all of the elements because of its high ionization energy, $\sim$25~eV, and basically zero electron affinity.\cite{Moore_1971, Hotop_1985_JPhysChemRefData}

At atmospheric conditions, helium is a constituent of a number of cationic and anionic complexes,\cite{Hotokka_1984_MolecularPhysics,Koch_1987_JACS,Frenking_1989_JPhysChem,Wilson2002,Li2005} including HHe$^+$, already characterized in 1925. \cite{Hogness1925} It has also been trapped in fullerenes \cite{Saunders1994,Saunders1996} and experiments suggest that other weak interactions besides van der Waals (vdW) are important for the formation of the CCl$_4$-He complex~\cite{Cappelletti2015}. However, a compound where helium forms a genuine chemical bond in an electrically neutral molecule has yet to be experimentally discovered. First principles calculations have proposed a number of potentially synthesizable helium-containing vdW complexes including HeBeO, HeLiF, HeLiH, and HeBN \cite{Frenking:1988a,Koch_1987_JACS,Borocci2006,Borocci2020}, where small highly-polar molecules perturb the noble gas atom. Metastable complexes such as (HeO)(LiF)$_2$, which consists of a covalently bonded He-O molecule placed within an LiF ferroelectric cavity \cite{Grochala:2012a}, and HHeF\cite{Wong:2000a}, where  a non-negligible amount of charge is transferred from helium, have also been proposed. Another example where chemical bonding, this time between two helium atoms, was suggested is in hypothetical He$_2$@B$_{12}$N$_{12}$ wherein the confinement within the small molecular cage coerces the noble gas atoms to interact~\cite{Saha2019}. 

With the notable exceptions of  He@C$_{60}$ and He@C$_{70}$~\cite{Morinaka2013}, helium has not been incorporated into a solid crystalline material at ambient conditions.  Because high-pressures can be harnessed to create compounds with novel compositions and electronic structures,  making use of unique bonding schemes~\cite{Grochala2007,Miao2020Nature,Hilleke2023}, materials and planetary scientists have searched for stable or metastable helium-containing compounds under pressure for decades.\cite{Miao2017Nature,Miao_2020_FiC,Grochala_2018_FoundChem}. In this context, crystal structure prediction (CSP) techniques\cite{Zurek2015,Zurek2017} combined with first-principles calculations have uncovered intriguing materials that may be constituents of ice-giant planets including  (H$_2$O)$_2$He at 300 GPa \cite{Liu_2015_PhysRevB}, various He-NH$_3$ \cite{Liu_2020_PhysRevX,Shi_2020_NatureComm} and He-H$_2$O \cite{Liu:2019a} phases that possess superionic states at high-temperatures, and He-CH$_4$ combinations, where temperature induces a ``plastic'' behavior characterized by protons that rotate about a fixed carbon atom in methane~\cite{Gao_2020_NationalSciRev}.  Another compound predicted to be superionic at high temperatures, FeHe, and a related FeHe$_2$ phase both become stable at pressures typical of iron-core white dwarf stars, between 4 and 12 TPa \cite{Monserrat_2018_PhysRevLett}. Moreover, it was proposed that helium could be used as a template to synthesize novel allotropes of nitrogen via quenching phases that are stable at high pressure -- including HeN$_4$  \cite{Li_2018_NatureComm}, HeN$_{22}$ \cite{Hou_2021_PhysRevB} and HeN$_3$ \cite{Wei_2019_JAlloy} -- to atmospheric conditions. 

A good example of the convergence of computation and experiment was the CSP-aided prediction of a stable Na$_2$He compound that was synthesized above 113~GPa \cite{Dong_2017_NatureChem}. A follow-up theoretical investigation explained why this crystal lattice is stable though no genuine chemical bonds are formed between Na and He.\cite{Liu_2018_NatureComm} Na$_2$He is a high pressure electride, consisting of cationic sodium cores and anionic electron pairs trapped in interstitial sites, so its formula may be written as [(Na$^+$)$_2e^{2-}$]He. First-principles calculations illustrated that for an ionic compound with a 2:1 or 1:2 ratio of cations to anions, the helium atoms can be placed between the two ions of the same charge thereby decreasing the repulsion between them and lowering the resulting Madelung energy.\cite{Liu_2018_NatureComm} Thus, the role of helium within [(Na$^+$)$_2e^{2-}$]He is to mitigate the electrostatic repulsion between nearby Na$^+$ ions, stabilizing the compound without the necessity of forming chemical bonds. The [(Na$^+$)$_2e^{2-}$] sublattice of Na$_2$He can as well be described as that of the anti-fluorite structure, where the He atoms and the interstitial electrons lie on the FCC sites, while the sodium atoms occupy the tetrahedral sites.  CSP calculations have uncovered other helium-containing systems that are stabilized via the aforementioned mechanism, many that assume similar structure types. This includes FeO$_2$He -- where the He and Fe atoms sit on the face-centered cubic (FCC) sites of a fluorite lattice -- stable at $P/T$ conditions relevant to the core-mantle boundary~\cite{Zhang_2018_PhysRevLett}, as well as compounds that adopt anti-fluorite lattices including HeK$_2$O, HeNa$_2$O, HeK$_2$S and HeRb$_2$S~\cite{Gao_2019_PhysRevMAt}. Moreover,  helium mixed with the electrostatic molecules water and ammonia are also predicted to be stable when squeezed.~\cite{Bai_2019_CommChem} 

The remaining helium-bearing compounds that have been synthesized were made at significantly lower pressures than Na$_2$He, typically $<$10~GPa. This includes the formation of solid helium-nitrogen vdW compounds~\cite{Vos_1992_Nature}, and the adsorption of the small helium atom -- sometimes serendipitously because of its use as a pressure transmitting medium~\cite{Zurek:2017f} that retains hydrostaticity until $\sim$40~GPa~\cite{Klotz2009}-- into water-ice clathrate cages~\cite{Londono_1988_Nature,Londono_1992_JChemPhys}, and  porous compounds such as SiO$_2$ \cite{Yagi_2007_PhysRevB,Shen_2011_PNAS,Sato2011} and As$_4$O$_6$.\cite{Gunka_2015_CGaD,Sans_2016_PhysRevB}. The proclivity of helium incorporation into materials that contain hollow interstices was leveraged in the synthesis of the first helium-containing perovskite, [He]$_2$[CaZr]F$_6$, which was proposed as a gas storage material~\cite{Hester_2017_JACS,Lloyd_2021_ChemofMat}. [CaZr]F$_6$ is a neutral hybrid double perovskite possessing vacant A-sites, and so, it is naturally porous. Helium could be inserted into the A-site cavities using mild pressures, with a solubility higher than in silica-glass or in cristobalite. Helium incorporation was shown to affect the mechanical properties of this perovskite, altering its compressibility and delaying pressure-induced amorphization~\cite{Hester_2017_JACS,Lloyd_2021_ChemofMat}. 

Herein, we present CSP-guided first-principles calculations to explore the incorporation of helium into vacant A-site perovskites with the MX$_3$ stoichiometry that are analogous to [CaZr]F$_6$.~\cite{Hester_2017_JACS,Lloyd_2021_ChemofMat} In particular AlF$_3$, GaF$_3$, InF$_3$, ScF$_3$ and ReO$_3$, were selected because their  pore sizes may be large enough to host helium, but not too large to let it move freely within the cavities. The most stable structures were found up to mild pressures of $\sim$10~GPa, focusing in particular on the topology of the pores found within them and the incorporation of helium into these pores. The thermodynamics associated with the helium-insertion mechanism were computed, to guide the development of future synthesis pathways. A detailed investigation of the preferred structures, band-gaps, bulk moduli and chemical bonding is performed. The insertion of helium, in particular into the perovskites with the smaller pores, greatly enhances their bulk moduli, presenting helium as a ``not so innocent'' guest. Though we do not find helium to form any ionic or covalent chemical bonds with the perovskite framework, signatures of weak interactions are calculated for compounds that contain small pores.

\section{Methods}
The open-source evolutionary algorithm (EA) \textsc{XtalOpt}~\cite{Lonie2011, Falls_2020_JPhysChemC}, version 12.1,\cite{Zurek:2018j} was employed to predict the most stable structures for the MX$_3$ and HeMX$_3$ stoichiometry  (M = Al, Ga, In, Sc and Re; X = F and O) phases. 
The initial generation consisted of random symmetric structures that were created by the \textsc{RandSpg} algorithm~\cite{Avery2017}, together with known $R\bar{3}c$ VF$_3$-type and $Pm\bar{3}m$ ReO$_3$-type frameworks that were added as initial seeds.  A sum of the atomic radii scaled by a factor of 0.6 was used to determine the shortest allowed distances between pairs of atoms. The \textsc{XtalComp} algorithm \cite{Lonie2012} was used to identify and remove duplicate structures from the breeding pool. CSP runs were performed on cells with 2-9 formula units at 1 and 10~GPa. The lowest enthalpy structures were relaxed from 0-10~GPa, and their relative enthalpies and equations of states are given in the Supporting Information.

Geometry optimizations and electronic structure calculations were performed using the Vienna Ab Initio Simulation Package (VASP), version 5.4.1~\cite{Kresse_1994_PhysRevB,Kresse_1999_PhysRevB}. The PBE~\cite{Perdew1996,Perdew1996a} exchange-correlation functional was employed in the CSP searches and the optB88-vdW functional~\cite{Dion_2004__PhysRevLett,Klimes_2010_JPhysCondMatt,Klimes_2011_PhysRevB},  which accounts for dispersion interactions, was used for the final relaxation of the lowest enthalpy structures. The projector augmented wave (PAW) method~\cite{Blochl_1994_PhysRevB} was used to treat the core states in combination with a plane-wave basis set with an energy cutoff of 650~eV (1100~eV)  for the CSP (final) calculations. The He 1s$^2$, F 2s$^2$2p$^5$, Al 3s$^2$3p$^1$, Ga 3d$^1$$^0$4s$^2$4p$^1$, In 4d$^1$$^0$5s$^2$5p$^1$, Sc 3s$^2$3p$^6$3d$^1$4s$^2$, O 2s$^2$2p$^4$ and Re 5p$^6$5d$^5$6s$^2$ states were treated explicitly (Table S1). The $k$-point meshes were generated using the $\Gamma$-centered Monkhorst-Pack scheme~\cite{Monkhorst1977} and the number of divisions along each reciprocal lattice vector was selected so that the product of this number with the real lattice constant was greater than or equal to 50~\AA{}. Dynamic stability was determined via phonon calculations in the harmonic approximation performed with the Phonopy package~\cite{Togo_2008_PhysRevB}, on supercells whose lattice vectors measured at least 10~\AA{}. The enthalpies of formation of various HeMF$_3$ phases were calculated as $\Delta H = H\text{(HeMX$_3$)} -$ ($H\text{(MX$_3)$}+H\text{(He)}$). Experiments performed on helium at temperatures close to 0~K suggest the hexagonal close packed (hcp) phase is preferred over the face centered cubic (fcc) structure~\cite{Loubeyre_1993_PhysRevLettb}. These two phases were isoenergetic within the optB88-vdW functional, consequently, we opted to use the hcp phase of helium as the reference state.

Analyzing the extreme, or nodes, of the radial function obtained via Hartree-Fock calculations that used a minimal basis set of Slater-type orbitals Clementi \emph{et al.}~\cite{Clementi1967} proposed an atomic radius of 0.31~\AA, yielding a sphere volume of 0.125~\AA{$^3$}. Therefore, the volumes of the pores in the A-site vacant perovskites were estimated using the Mercury 4.0 package~\cite{MacRae2020} with a probe of radius 0.31~{\AA} -- an approximation of the helium atom -- and a grid spacing of 0.1~{\AA}. The probe radius was moved around the cavity until it touched the vdW radius of an atom comprising the perovskite, yielding a pore volume that is accessible to the helium guest. For the HeMX$_3$ systems, the volume of the pores were obtained with the same procedure after manually removing the helium atoms from the structures. We note that many ways have been devised to define radii, of various types, associated with atoms. Turning to the ambient-pressure vdW radius of He, Rahm \emph{et al.}~\cite{Rahm2016a} defined this via a cutoff of 0.001~e/Bohr$^3$ on the DFT calculated electron density (at the PBE0/ANO-RCC level of theory) yielding a value of 1.34~\AA{} and a volume of 10.1~\AA{$^3$}, which is very close to the vdW radius obtained by Alvarez~\cite{Alvarez2013} (1.43~\AA) and Bondi~\cite{Bondi1964a} (1.40~\AA) from crystallographic data. The vdW radii of He as a function of pressure that are presented were calculated by Rahm \emph{et al.}~\cite{Rahm2020} using the eXtreme Pressure Polarizable Continuum Model~\cite{Cammi2015} developed by Cammi and downloaded from the SHARC - Atoms Under Pressure webpage \cite{sharc}. 

To better understand the chemical bonding we calculated the crystal orbital overlap/Hamilton populations (COOP~\cite{Hughbanks_1983_JACS}/COHP~\cite{Dronskowski_1993_JPhysChem}) and the negative of the COHP integrated to the Fermi level (-ICOHPs)~\cite{Deringer_2011_JChemPhysA} using the LOBSTER package~\cite{Nelson_2020_JCompChem}. Charge-spilling values were calculated to be <~1$\%$, except for HeReO$_3$, where the charge spilling value was $\sim$1.5$\%$.
Several types of atomic charges were calculated. Bader charges were obtained either with the BADER code~\cite{Tang2009}, if from VASP outputs, of directly with the periodic  BAND code~\cite{Philipsen_2022_BAND}. BAND was also employed to calculate Hirshfeld and Mulliken charges~\cite{FonsecaGuerra2004} at the BP86-D4/TZ2P~\cite{Perdew_1986_PhysRevB,Becke_1988_PhysRevA,Caldeweyher_2020_PCCP} level of theory  on the VASP optimized geometry. 
Finally, LOBSTER~\cite{Nelson_2020_JCompChem} was also used to evaluate Mulliken and L\"{o}wdin charges from VASP outputs.

Mechanical properties including the bulk modulus, $B$, and shear modulus, $G$, were calculated using the AFLOW Automatic Elastic Library~\cite{Toher_2017_PhysRevMat}. The Vickers hardness was estimated using the Teter model, $H_\text{v,Teter}=0.151G$~\cite{Teter1998}. It has been shown that the Teter model in combination with DFT-calculated shear moduli give reliable estimates of $H_\text{v}$ for a broad range of materials~\cite{Zurek:2019b}.

\section{Results and discussion}
\subsection{Intercalating He into [CaZr]F$_6$}

Before proceeding to our theoretical investigation of the insertion of helium into A-site vacant perovskites, we investigated stuffing helium into [CaZr]F$_6$, since this process has been observed and studied experimentally~\cite{Lloyd_2021_ChemofMat}. The $a$-lattice parameter for this perovskite, whose crystal lattice is noticeably porous, was measured to be \emph{ca.}\ 8.51~\AA{} at 10~K, contracting  to 8.48~\AA~at ambient conditions due to the large negative thermal expansion that characterizes [CaZr]F$_6$~\cite{Hancock2015}. The PBE functional yielded a slightly too-large lattice parameter of 8.61~\AA{} (at 0~K), whereas the value obtained with the optB88-vdW functional, 8.56~\AA{}, which treats vdW interactions self-consistently, was closer to experiment. It is not surprising that pressure causes this porous structure to undergo amorphization at a mere 0.5~GPa when neon is used as the pressure transmitting medium~\cite{Lloyd_2021_ChemofMat}.

\begin{figure}[H]
	\centering
		\includegraphics[width=\figurewidth]{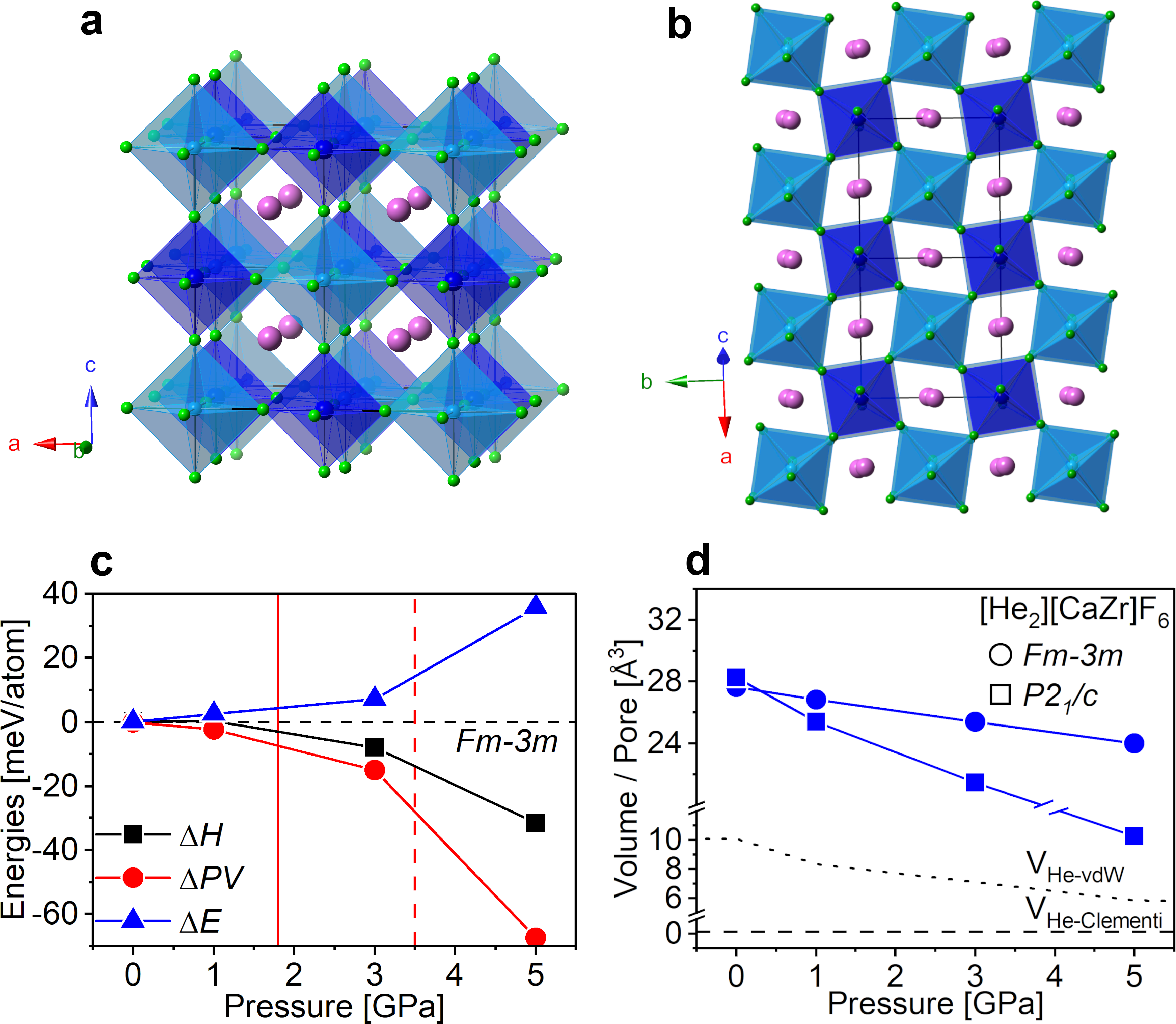}
	\caption{Illustration of the (a) $Fm\bar{3}m$ [He]$_2$[CaZr]F$_6$ phase (isomorphic with $Fm\bar{3}m$ [CaZr]F$_6$), and the (b)  $P2_{1}/c$ [He]$_2$[CaZr]F$_6$ phase with the green/blue/pink balls representing F/Ca,Zr/He atoms. (c) The relative enthalpies ($\Delta H$) between $Fm\bar{3}m$ and $P2_{1}/c$ symmetry [He]$_2$[CaZr]F$_6$ phases and the energetic ($\Delta E$) and pressure-volume ($\Delta PV$) contributions towards them as a function of pressure. The red dashed line marks the pressure at which [He]$_2$[CaZr]F$_6$ begins decomposing in experiments~\cite{Lloyd_2021_ChemofMat}, and the red solid line the pressure at which the  $Fm\bar{3}m$ was observed to undergo a phase transition~\cite{Lloyd_2021_ChemofMat}. (d) Volume of the pores within the $Fm\bar{3}m$ and $P2_1/c$ [He]$_2$[CaZr]F$_6$ phases as a function of pressure. The black horizontal dashed lines represent the volume of helium as estimated using the Clementi atomic radii and the vdW radii under pressure by Rahm-Cammi (see Methods for further information). }
	\label{fig:He2CaZrF6}
\end{figure}

When helium is used as the pressure transmitting medium instead, the amorphization is delayed until $\sim$3.5~GPa, aided by an uptake of helium into the pores and accompanied by a phase transition of $Fm\bar{3}m$ [He]$_2$[CaZr]F$_6$ (Figure \ref{fig:He2CaZrF6}(a)) to an unknown tetragonal space group (suggested to be either $I4/m$ or $P4/mnc$) above 1.8~GPa and at low temperature~\cite{Lloyd_2021_ChemofMat}. Our evolutionary structure searches found the most stable phase at 1~GPa possesses the monoclinic $P2_{1}/c$ spacegroup instead, and subsequent calculations showed that it was more stable than the proposed $I4/m$ and $P4/mnc$ phases (Figure S1). In the $Fm\bar{3}m \rightarrow P2_1/c$ transformation, the [CaZr]F$_6$  octahedra rotate in a way that resembles what would be expected for a transition to the $P4/mnc$ spacegroup, with the main difference being an additional tilting along the $c$-direction that induces a staggered, rather than a linear, orientation of the He atoms within the 1-dimensional channels, as shown in  Figure\ \ref{fig:He2CaZrF6}(b).
The staggering allows $P2_{1}/c$ to assume a slightly smaller volume than the previously proposed tetragonal phases (285.4~\AA$^3${} for $P2_{1}/c$, 285.7~\AA$^3${} for $P4/mnc$ and 285.8~\AA$^3${}  for $I4/m$). In good agreement with experiment, we predict the  $Fm\bar{3}m\rightarrow P2_{1}/c$ transition to occur just above 1~GPa (Figure\ \ref{fig:He2CaZrF6}(c)). Under pressure, the enthalpy, which is a sum of the energetic and pressure/volume terms, $H = E + PV$, is the key thermodynamic variable determining phase stability. The $Fm\bar{3}m \rightarrow P2_1/c$ structural distortion is accompanied by an increase in internal energy that is overshadowed by a decrease in the $PV$ term, which is the driving force for this phase transition above 1~GPa.

The cavities within [CaZr]F$_6$ are quite large, measuring 27.5~\AA{}$^3$ at zero pressure within our optB88-vdW calculations. This value is more than double the estimated vdW volume of helium, which can easily fit into each pore. Therefore, the ambient pressure lattice parameter of 8.564~\AA{} does not change much upon helium insertion, increasing by only 0.006~\AA. Helium incorporation has a negligible effect on the size of the pore, whose volume expands to 27.6~\AA{$^3$} (Figure\ \ref{fig:He2CaZrF6}(d)). The uptake of helium is calculated to be favorable by 11.0~meV/atom within the optB88-vdW functional at zero pressure. Attractive dispersion interactions between the encapsulated helium atom and the perovskite framework are the likely culprit of the exothermic reaction, 
which is slightly endothermic (by 1.4~meV/atom) within the PBE functional (Table S2). Helium incorporation has a negligible influence on the band gap of the perovskite,  which was calculated to be $\sim$6.9~eV with the nonlocal optB88-vdW functional (Table S3). The bulk modulus was computed to increase from 58.6 to 61.6~GPa upon helium insertion, which is a somewhat smaller  than the expansion that has been measured experimentally ($\sim$36$\rightarrow$47~GPa~\cite{Hancock2015,Lloyd_2021_ChemofMat}).  

As pressure increases the volume of the pores within the $P2_{1}/c$ [He]$_2$[CaZr]F$_6$ phase becomes significantly smaller than within the $Fm\bar{3}m$ polymorph (Figure\ \ref{fig:He2CaZrF6}(d)). At $P>3$~GPa, the calculated volume of each pore within the monoclinic phase abruptly collapses to 10.3~\AA{}$^3$. Notably, in experiments [He]$_2$[CaZr]F$_6$ begins to amorphize at about this pressure, and this process is completed by 6~GPa~\cite{Lloyd_2021_ChemofMat}. Our calculated enthalpic trends show that $P2_{1}/c$ [He]$_2$[CaZr]F$_6$ is preferred up to at least 5~GPa from the constituent perovskite and noble gas atom. Nonetheless, the abrupt shrinking of the cavities signals a severe distortion of the framework, which would likely be the cause of the incipient amorphization of [He]$_2$[CaZr]F$_6$ at higher pressures.

\subsection{Stuffing Helium into A-site Vacant Perovskites}

\subsubsection{Quantifying the Pore Sizes}
Let us introduce the MX$_3$ perovskites we considered in this study, focusing on their structures and the geometric peculiarities of their important vacant A-sites where the helium atom may be stuffed. AlF$_3$, GaF$_3$ and InF$_3$ belong to the VF$_3$-type materials family crystallizing in the $R\bar{3}c$ space group~\cite{Hoppe_1984_JFluoChem,Jorgensen_2010_HighPresRes}. Experiments revealed this rhombohedral structure remains the ground state up to 50~GPa for AlF$_3$\cite{Stavrou_2015_JChemPhys} and 28~GPa for GaF$_3$~\cite{Jorgensen_2010_HighPresRes}. Comparison of the enthalpy calculated for $R\bar{3}c$-AlF$_3$ with the enthalpies of other known stable and metastable polymorphs that can be made at high temperatures or using molecular templates~\cite{Konig_2010_JFluoChem, Krahl_2017_CatSCiandTech} revealed that it was the preferred geometry at both 0 and 10~GPa (Table S4). In addition to this series of VF$_3$-type perovskites we also considered ScF$_3$ and ReO$_3$, because these are examples of A-site vacant perovskites that contain transition metals that do not possess a magnetic moment.  At ambient conditions, both compounds adopt the $Pm\bar{3}m$ space group~\cite{Jorgensen2004a, Greve_2010_JACS}, which is isomorphic to the $Fm\bar{3}m$ [CaZr]F$_6$ structure. ScF$_3$ transitions to the rhombohedral $R\bar{3c}$ space group  at 0.7~GPa \cite{Aleksandrov_2002_ExpTheoPhys}, while ReO$_3$ transforms to the $Im\bar{3}$ space group at \emph{ca.}\ 1.3~GPa, which remains stable up to 13.2~GPa, where it transforms once again to a rhombohedral phase, stable up to 52 GPa~\cite{Jorgensen2000,Jorgensen2004a}.

\begin{figure}[H]
	\centering
		\includegraphics[width=0.7\figurewidth]{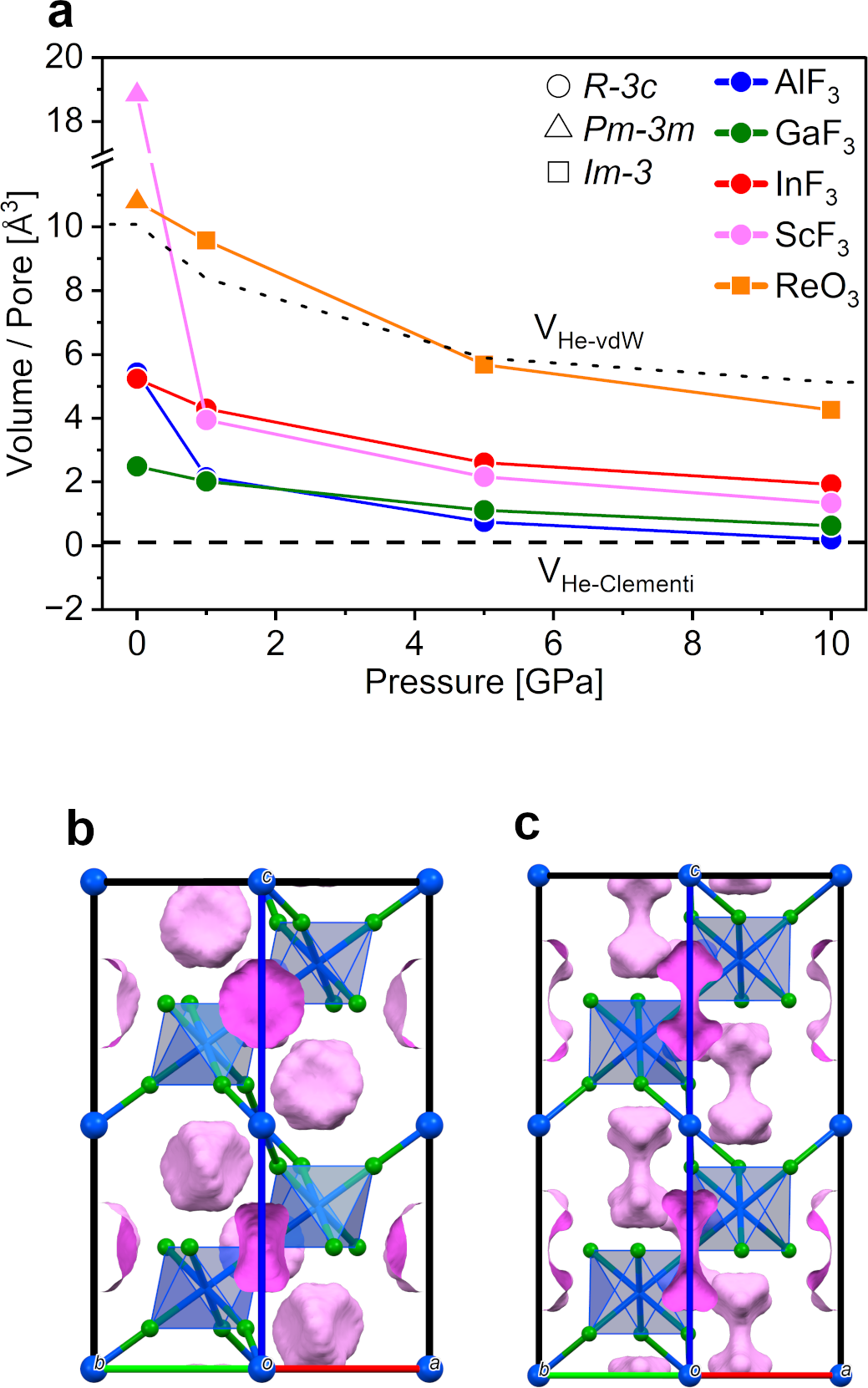}
	\caption{(a) Volume of the pores in MX$_3$ compounds (color code represents chemical composition, and symbol-shape the space group) under pressure. Illustrations of the pores (pink) accessible to the helium guest atoms within $R\bar{3}c$ symmetry (b) AlF$_3$ and (c) InF$_3$. Spherical pores similar to those shown in (b) are characteristic of AlF$_3$, ScF$_3$ and ReO$_3$, whereas figure eight shaped pores resembling those shown in (c) are found within GaF$_3$ and InF$_3$ at 0~GPa. Green/blue balls represent M/X atoms (M = Al, Ga, In, Sc, Re; X = O, F).}
	\label{fig:V_pores_shape}	
\end{figure}

Unsurprisingly, the volumes that can be ascribed to the vacant A-sites in these MF$_3$ (M~=~Al, Ga, In, Sc) and ReO$_3$ compounds (Figure\ \ref{fig:V_pores_shape}(a)) turn out to be significantly smaller than in the [CaZr]F$_6$ perovskite at 1~atm, in-line with their persistence when squeezed (so far their pressure-induced decomposition has not been observed). 
With increasing pressure the volumes of the voids decrease, sometimes quite precipitously, to minimize the $PV$ contribution to the enthalpy. This behavior coincides with the first rule  that characterizes the response of molecular crystals to modest pressures of $\sim$10~GPa that was proposed by Grochala and co-workers: ``vdW space is most easily compressed''~\cite{Grochala2007}. The volumes of the pores within ScF$_3$ at 0~GPa and ReO$_3$ up to $\sim$5~GPa are larger than the estimated vdW volume of a single helium atom, suggesting that the $PV$ term to the enthalpy will readily favor helium incorporation into these phases. For the VF$_3$ family, the pore sizes are markedly smaller than the vdW radius of helium, hinting that the incorporation of the inert noble gas element may not be favorable, unless other factors help to stabilize the He-intercalated compound. 

Does the identity of the perovskite affect the topology of the pores?  Our analysis revealed two basic cavity shapes -- one that is spherical and one that is shaped like a figure eight -- as shown in Figure\ \ref{fig:V_pores_shape}(b) and (c), respectively. At ambient pressure, AlF$_3$, ScF$_3$ and ReO$_3$ present spherical shapes, meanwhile, GaF$_3$ and InF$_3$ are characterized by cavities with a prominent figure eight shape.

Despite the fact that AlF$_3$, GaF$_3$ and InF$_3$ crystallize in the rhombohedral $R\bar{3c}$ space group, the shapes of the pores within them are quite different. As expected, the M-F distances increase in going down the group and are measured to be 1.81, 1.91 and 2.10~\AA{}, respectively, in the 0~GPa optimized compounds. At the same time as these bonds elongate, the M-F-M angles within the octahedra become more acute measuring 157.8, 139.5 and 134.7~$^{\circ}$ within AlF$_3$, GaF$_3$ and InF$_3$, respectively, implying a larger degree of rotation of the octahedra for Ga and In, yielding the figure eight cavities.  
This type of packing is a natural structural relaxation allowed by the vacancy in the A-site, and favored when the M-F distance is longer, like in the case of GaF$_3$ and InF$_3$. 
For example, despite the longer M-F bond lengths in InF$_3$ as compared to AlF$_3$, at zero pressure the two compounds possess very similar cavity volumes (Figure\ \ref{fig:V_pores_shape}(a)). Nonetheless within AlF$_3$ this corresponds to a volume that is spherically accessible, meanwhile for InF$_3$, this value technically refers to the sum of the volumes of the two lobes (Figure\ \ref{fig:V_pores_shape}(c)). Consequently, if the figure eight type of packing is favored already at ambient conditions, shrinking the pores in this eight shape, will already hinder the intake of helium. 

\subsubsection{Crystalline HeMX$_3$ Structures}
\begin{figure} [H]
	\centering
		\includegraphics[width=1\figurewidth]{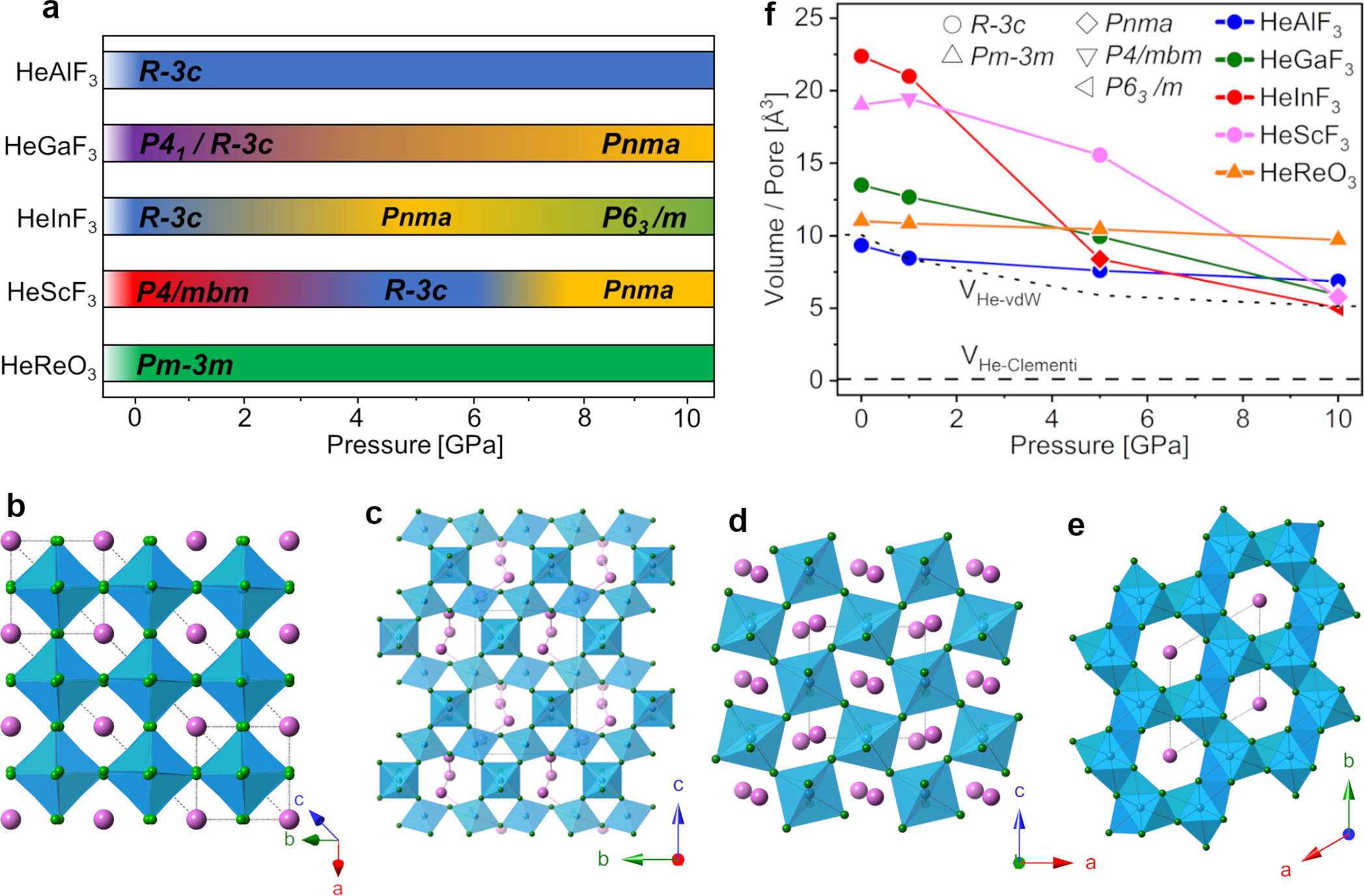}
	\caption{(a) Computed phase diagrams of the MX$_3$He compounds. Illustrations of the predicted structures: (b) $R\bar{3}c$ (HeAlF$_3$, HeGaF$_3$, HeInF$_3$ and HeScF$_3$), (c) $P4_{1}$ (HeGaF$_3$),  (d) $Pnma$ (HeGaF$_3$, HeInF$_3$, HeScF$_3$), and (e) HeInF$_3$ $P6_{3}/m$. (f) Volume of the pores as a function of pressure calculated for the HeMX$_3$ species (color code represents composition, and symbol-shape the space group).}
	\label{fig:phases}
\end{figure}

To determine what structures might be adopted when helium is incorporated into the five A-site vacant perovskites considered in this study, evolutionary crystal structure searches, seeded with the known VF$_3$-type and $Pm\bar{3}m$ symmetry frameworks, which were stuffed with helium atoms, were performed at 1 and 10~GPa. The most promising structures found were relaxed within this pressure range and their relative static enthalpies, and finite temperature additions towards them, were compared, and will soon be discussed. Figure \ref{fig:phases}(a) presents a phase diagram approximately illustrating the pressure ranges where these particular structure types were preferred (based on the static lattice enthalpies and the dynamic stability alone). Since the phonons of these phases were computed in increments of 2~GPa, the borders delineating a structure's domain of stability are not drawn abruptly, and are purposefully fuzzy. Figures \ref{fig:phases}(b-e) illustrate the main structure types that were found, focussing on the pores that house the helium atoms. Similar to what was observed for [CaZr]F$_6$, the helium-stuffed perovskites often differ in important ways from their vacant A-site analogues. In fact, with the exception of the HeAlF$_3$ and HeReO$_3$ compounds, all of the compositions studied here are predicted to undergo a series of phase transitions between 0-10~GPa. 
Inserting helium into the most stable ambient pressure AlF$_3$ polymorph does not affect the symmetry of this perovskite, likely because of the spherical pore shape found within it. In fact, the resulting $R\bar{3}c$-HeAlF$_3$ structure (Figure\ \ref{fig:phases}(b)) remains the most stable He@AlF$_3$ geometry up to at least 10~GPa. Helium incorporation increases the too-small pore size so that it is at least as large as the vdW volume of this noble gas atom within most of the pressure range studied (Figure \ref{fig:phases}(f)). 

The figure eight type pore in $R\bar{3}c$-GaF$_3$ does not possess a shape within which helium would easily fit  (Figure\ \ref{fig:V_pores_shape}c). Our calculations found two HeGaF$_3$ phases that were dynamically stable and nearly isoenthalpic at 1~GPa. The first of these, with $P4_1$ symmetry  (Figure \ref{fig:phases}(c)), differs significantly from its parent perovskite structure, suggesting that large barriers may hinder its formation. Nonetheless, we discuss the structural peculiarities of this phase, as it still might be synthesized. This tetragonal structure contains two sets of GaF$_6$ octahedra, one whose apical fluorine atoms point along the $b$ crystallographic axis and ones that are rotated so that their apical fluorines are oriented along the $a$ crystallographic axis. These two sets of octahedra are joined by vertex-sharing equatorial fluorine atoms creating a framework characterized by channels extending along the $a$ and $b$ crystallographic directions, and connected by a 1-dimensional helical channel along the $c$ direction that is plenty large enough to be filled with helium atoms whose He-He distances measure 2.31 and 2.70~\AA{}. At higher pressures the previously discussed $R\bar{3}c$ symmetry structure predicted for HeAlF$_3$ is the only one that is preferred. At 1~GPa the M-F-M angles within GaF$_3$, which are related to the rotation of the adjacent octahedra (a$^-$a$^-$a$^-$ tilt in Glazer notation~\cite{Glazer1972}), increase from 138.1$^{\circ}$ to 169.4$^{\circ}$ upon helium insertion, which is similar to the 173.1$^{\circ}$ calculated for HeAlF$_3$ at the same pressure. Therefore, the incorporation of helium distorts the perovskite framework so that the figure eight shaped pores, containing two small cavities, merge into a single larger pore within which the noble gas atom can fit. Consequently, the pore size within $R\bar{3}c$-HeGaF$_3$ is larger than within $R\bar{3}c$-HeAlF$_3$ (Figure \ref{fig:phases}(f)) thanks to its longer M-F bonds and because the presence of helium atoms within the pores hinders the rotations and other distortions of the surrounding octahedra.  

At a pressure close to 10~GPa, HeGaF$_3$ transforms into the orthorhombic $Pnma$ space group (Figure\ \ref{fig:phases}(d)). This configuration, which is also predicted for HeInF$_3$ and HeScF$_3$ under pressure, is commonly known as the GdFeO$_3$ structure-type~\cite{Stoumpos_2015_AccChemRes,Akkerman_2020_ACSEnergyLett}, and it is the result of an a$^+$b$^-$b$^-$ tilting~\cite{Glazer1972}. In this geometry, the atoms of helium are displaced with respect to the center of the cavity that runs along the $b$-axis, periodically alternating between its two opposing sides (Figure\ \ref{fig:phases}(d)). The transition to this phase is a direct effect of the figure eight shaped cavities discussed above. Compression of the MX$_3$ framework containing this cavity shape accentuates its asphericity, so it eventually split into two lobes, thereby forcing the noble gas to move from the center of the cavity. The unoccupied lobes are then used by the framework to further relax, triggering a concomitant tilting of the octahedra and reducing the volume of the pores (\emph{c.f.} the volumes ascribed to the pores in $R\bar{3}c$ vs.\ $Pnma$ HeInF$_3$ and $P4/mbm$  vs.\ $Pnma$ HeScF$_3$ shown in Figure\ \ref{fig:phases}(f)), and generating the alternating pattern of helium atoms visible in Figure\ \ref{fig:phases}(d).

The $R\bar{3}c$ and $Pnma$ phases predicted to be the most stable for HeInF$_3$ were isotypic with those described above for HeGaF$_3$, except the structural transition was computed to occur at a slightly lower pressure for the heavier triel. A $P4_1$ symmetry analogue, however, was not found in our evolutionary runs and calculations showed it was thermodynamically and dynamically unstable at 1~GPa. Incorporation of helium yielding the $R\bar{3}c$ phase results in an increase of the In-F-In angles from  133.2$^{\circ}$ to 159.3$^{\circ}$ at 1~GPa. In both the empty and stuffed $R\bar{3}c$ InF$_3$ phases the degree of rotation of the octahedra is more pronounced than in the analogous $R\bar{3}c$ GaF$_3$ structures, likely because of the larger size of the metal cation and the longer M-F distances. As a result the pore size in HeInF$_3$ is more than double that found within HeAlF$_3$ up to 1~GPa, easily accommodating helium. The pores collapse as a consequence of the transition to the $Pnma$ phase (Figure \ref{fig:phases}(f)).    
By 10~GPa $P6_{3}/m$ HeInF$_3$ (Figure\ \ref{fig:phases}(e)) emerges as the most stable geometry. This phase is characterized by a 1-dimensional channel running along the $c$ crystallographic direction. Moreover, it contains 9-coordinated indium atoms at the center of a tricapped triangular prism of fluorine atoms, which resembles the ReH{$_9^{2-}$} anion. The transformation to this $P6_{3}/m$ geometry, not observed in HeGaF$_3$ under pressure, is facilitated in HeInF$_3$ by the larger radius of indium and its ability to form hypercoordinated bonds at lower pressures. This structural transition results in a substantial change of the topology and the local coordination environments within the structure. Nonetheless, the channels in this novel phase are just the right size to accommodate the vdW volume of helium at 10~GPa, suggesting that the noble gas could even be used as a templating agent to synthesize new MF$_3$ polymorphs -- a route that has been previously proposed as a technique to create new polymorphs of nitrogen~\cite{Li_2018_NatureComm,Wang2022a}.

The most stable HeScF$_3$ structure at 1~GPa assumes a tetragonal $P4/mbm$ geometry. This phase is not illustrated in Figure \ref{fig:phases} because it resembles the $Fm\bar{3}m$ [He]$_2$[CaZr]F$_6$ structure (which can be derived by doubling the unit cell of $Pm\bar{3}m$ ScF$_3$ that is preferred at 0~GPa and stuffing the voids with helium), except for the Sc-F-Sc angles in the $ab$ plane, which distort from a perfect 180$^\circ$ to 179$^\circ$, corresponding to a slight a$^0$a$^0$c$^+$ tilting. At 0~GPa, the pore size in $Pm\bar{3}m$ ScF$_3$ (Figure\ \ref{fig:V_pores_shape}(a)) is plenty big to accommodate the He atom, so no structural distortion is required. However, by 1~GPa the angles start to deviate from the ideal 180$^\circ$. At higher pressures this compound undergoes the same transformations observed in many of the previously discussed MF$_3$ systems: into $R\bar{3}c$ at $\sim$5~GPa, followed by a second transition to $Pnma$, which are each accompanied by a successive shrinking of the pore size until it is about the same as the vdW radius of helium.  

Finally, our calculations showed that the inclusion of helium within HeReO$_3$ acts as a stabilizer for the cubic $Pm\bar{3}m$ structure, evading the phase transition to the $Im\bar{3}$ space group observed for ReO$_3$ at \emph{ca.}\ 1.3~GPa. The pore size of ReO$_3$ shrinks from $\sim$11~\AA$^3${} to just over 4~\AA$^3${} in the $Im\bar{3}$ phase by 10~GPa (Figure\ \ref{fig:V_pores_shape}(a)). The $Pm\bar{3}m$ HeReO$_3$ phase, on the other hand, maintains quite a constant pore volume of about 11~\AA$^3${}, which is above the isolated vdW volume of He throughout the whole pressure range studied.

\subsubsection{Thermodynamics of Helium Absorption}
\begin{figure}[h!]
	\centering
		\includegraphics[width=0.5\figurewidth]{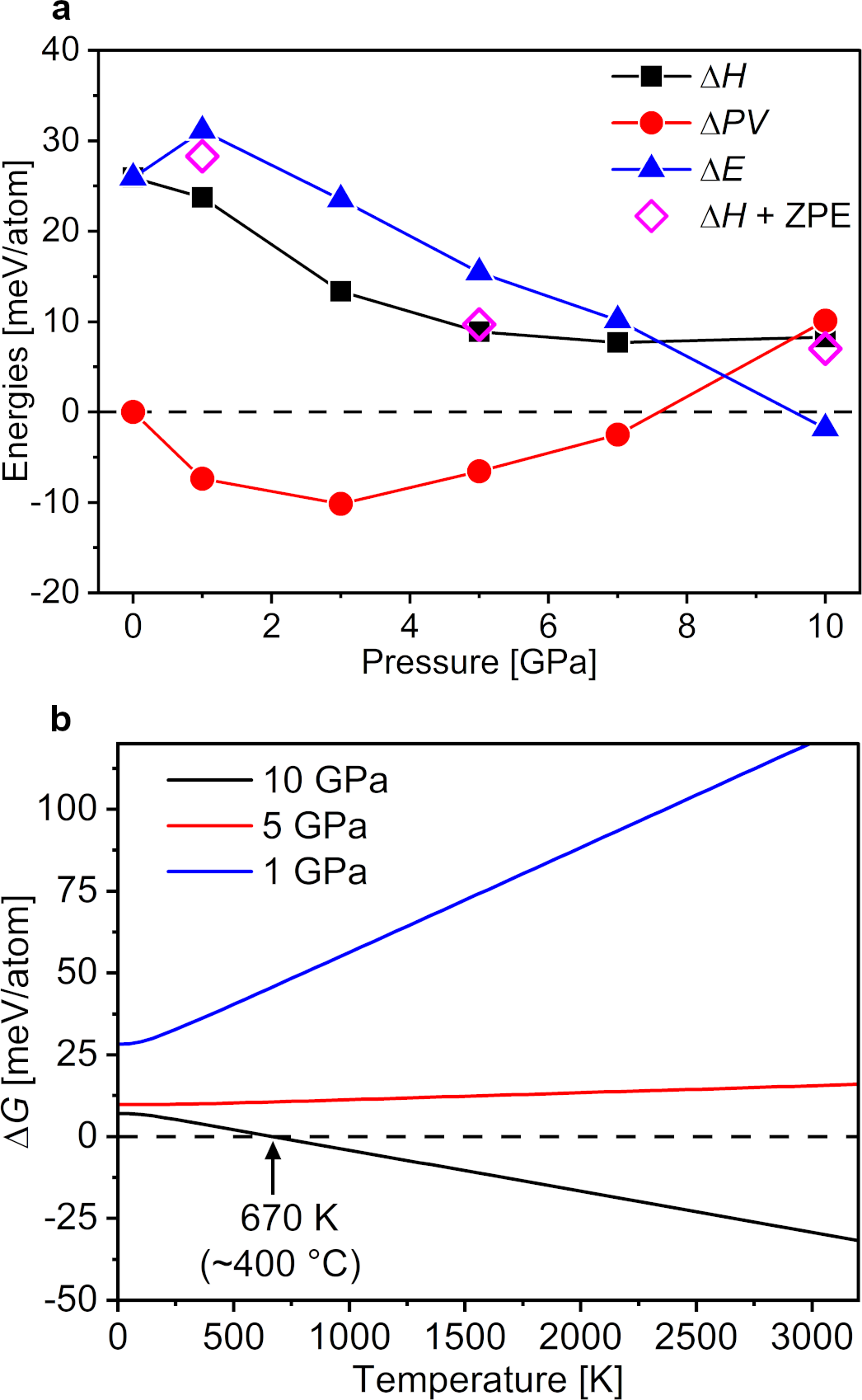}
	\caption{Thermodynamic terms associated with the reaction: $\text{AlF}_3+\text{He}\rightarrow \text{HeAlF}_3$. (a) The change in the enthalpy, $\Delta H$, as well as the pressure-volume, $\Delta PV$, and energetic, $\Delta E$, terms that comprise it as a function of pressure. The zero-point energy (ZPE) contribution to the static-lattice enthalpy is provided at select pressures. (b) The change in the Gibbs free energy, $\Delta G = \Delta H-T\Delta S$, at select pressures as a function of temperature. We remind the reader that  $k_\text{B}T\sim25.7$~meV at 298~K.}
	\label{fig:energy_terms}
\end{figure}

All of the predicted HeMX$_3$ compounds are dynamically stable in the range of pressures illustrated in Figure\ \ref{fig:phases}(a) (phonon band structures are provided in Figures S2-S6). The $P4_{1}$ phase of HeGaF$_3$, isoenthalpic with the $R\bar{3}c$ phase, was dynamically stable only at 1 GPa. Generally speaking, the change in the static-lattice enthalpy associated with helium incorporation ($\Delta H = H(\text{HeMX}_3)-H(\text{MX}_3)-H(\text{He})$) was calculated to be positive for most of the systems considered (Figure \ref{fig:energy_terms}(a) and Figure S7). Exceptions to this trend included  the formation of $P6_3/m$ HeInF$_3$ above $\sim$8~GPa, $P4/mbm$ HeScF$_3$ below $\sim$3~GPa, and HeReO$_3$ above 1~GPa. A data mining study of the DFT-computed 0~K energies of structures found in the Inorganic Crystal Structure Database concluded that $\sim$70~meV/atom corresponds to the 90$^{th}$ percentile of the DFT-calculated metastability~\cite{Sun2016}. According to this criterion, all of the dynamically stable phases reported here could potentially be made, provided an appropriate synthesis route can be found. 

Let us now consider how different terms contribute to the thermodynamics associated with helium incorporation.  Figure\ \ref{fig:energy_terms}(a) illustrates that $\Delta H$ for the formation of HeAlF$_3$ decreases as pressure is increased, reaching a minimum value of $\sim$8~meV/atom at 5~GPa, which remains nearly constant at higher pressures. Below 8~GPa the $PV$ term favors helium incorporation, as the volume of the pores approaches the atomic volume of helium (Figure \ref{fig:V_pores_shape}(a)). The magnitude of the energetic term, which is initially destabilizing, decreases under pressure so that by 10~GPa it favors the incorporation of the noble gas. Since helium is a light element, the zero-point energy (ZPE) arising from its vibrations might provide a non-negligible correction to the enthalpic change associated with its reaction. Previous computational studies have come to different conclusions on the importance of the ZPE: in reactions with ammonia~\cite{Shi_2020_NatureComm} and with alkali oxides and sulfides~\cite{Gao_2019_PhysRevMAt} the ZPE was generally found to be a destabilizing term; conversely, the ZPE stabilized water-helium compounds~\cite{Liu_2015_PhysRevB}. The effect of the ZPE on the enthalpy of formation of various helium containing materials might be dependent upon factors including the magnitude of the applied pressure, the atom-types and bonding that is involved in the reaction, and perhaps the level of theory used in the calculations.  For HeAlF$_3$ formation, the ZPE is destabilizing at 1~GPa (increasing the $\Delta H$ from 24.0 to 28.3~meV/atom), nearly negligible at 5~GPa, and very slightly stabilizing at 10~GPa  (decreasing $\Delta H$ from 8.4 to 7.0~meV/atom). 

Because high-pressure syntheses are typically aided by heating, sometimes to thousands-of-degrees Kelvin, we also calculated the change in the Gibbs free energy, $\Delta G$, associated with the formation of HeAlF$_3$, assuming the harmonic approximation (Figure\ \ref{fig:energy_terms}(b)). 

At low pressure (1~GPa) the entropic $T\Delta S$ contribution destabilizes HeAlF$_3$, in particular at high temperatures suggesting it is unlikely that this compound could form. Importantly, with increasing pressure this trend is reversed. At 5~GPa the $T\Delta S$ term is near zero and does not have an effect on the thermodynamics associated with helium incorporation. Notably, by 10~GPa the entropic contribution favors the formation of the ternary, suggesting that HeAlF$_3$ could be synthesized at this pressure above 670~K (provided it can enter the perovskite pores). 

Why is the trend in the Gibbs free energy profile inverted with increasing pressure, such that entropy favors the exsolution of helium from the ternary at 1~GPa, but its incorporation within the perovskite pores at 10~GPa? To understand this phenomenon, we must consider the large compressibility of pure helium. According to our DFT calculations, a single atom of helium in the hcp phase occupies 11.9~\AA$^3$ at 1~GPa, 8.4~\AA$^3$ by 5~GPa, and 7.0~\AA$^3$ by 10~GPa. Comparison of these atomic volumes with the volumes of the pores within the filled A-site perovskites (Figure \ref{fig:phases}(f)) shows that under pressure the enclathrated helium atoms have more room to move than within hcp helium. This decreased confinement of the helium atoms contained in the pores as compared to those in the elemental solid leads to a higher entropy. Indeed, comparison of the entropies of HeAlF$_3$, AlF$_3$ and He at 1~GPa with those at 10~GPa show that they all decrease with increasing pressure -- but the rate at which they do so is not the same, being nearly a factor of two greater for elemental helium.

Plots illustrating the variation of the enthalpy and Gibbs free energy associated with helium incorporation for the other vacant A-site perovskites studied are provided in Figure S7.  
For HeGaF$_3$, $\Delta H$ was always positive, hovering at $\sim$4~GPa for the most stable species predicted up to 10~GPa. The inclusion of helium within InF$_3$ became enthalpically favored by $\sim$8~GPa within the $P6_3/m$ phase, whereas for HeScF$_3$ it was preferred below $\sim$4~GPa. 
The uptake of helium within ReO$_3$ was favored above 1~GPa, with the magnitude of $\Delta H$ increasing with pressure to $\sim$-63~meV/atom by 10~GPa. The ZPE correction to the static lattice enthalpies did not have a considerable impact on the computed $\Delta H$ values ($\sim$1 meV, on average), sometimes increasing (\emph{e.g.} +1.2 meV/atom for\ $Pnma$ HeScF$_3$ at 10~GPa) and sometimes decreasing (\emph{e.g.}\  -0.5 meV/atom for\ $P4/mbm$ HeScF$_3$ at 10~GPa) them.  
The way in which the entropic contribution affected the Gibbs free energy varied greatly, and was dependent on the pressure and the geometry of the helium-intercalated perovskite, and also on the phase transitions of the host framework. 
Putting this all together, we found that the formation of the following phases was exergonic at the pressure and temperature conditions given in the braces: $R\bar{3}c$ HeAlF$_3$ (10~GPa, above 670~K), $R\bar{3}c$ HeGaF$_3$ (5~GPa, above 2300~K), $P6_{3}/m$ HeInF$_3$ (above 8 GPa, below 215~K), $P4/mbm$ HeScF$_3$ (1~GPa at all temperatures considered),  $R\bar{3}c$ HeScF$_3$ (below 4~GPa, then at 5~GPa above 200~K) and $Pm\bar{3}m$ HeReO$_3$ (above 1~GPa at most, if not all, temperatures). We hope these conditions are a useful guide for experimental groups interested in synthesizing these, or other helium bearing porous phases.

\subsubsection{Mechanical and Structural Properties}
It is easy to imagine why the A-site vacant perovskites, due to their porous structure, are often characterized by relatively small bulk moduli~\cite{Jorgensen_2010_HighPresRes}. It is also not a surprise that inclusion of helium in the vacant A-sites affects their mechanical properties, as has been observed for [CaZr]F$_6$ whose bulk modulus increased by $\sim$11~GPa upon helium intercalation \cite{Hester_2017_JACS, Lloyd_2021_ChemofMat}. Similarly, helium incorporation within silica glass resulted in a reduction of its compressibility so that $B$ increased from 14 to 56~GPa \cite{Yagi_2007_PhysRevB,Shen_2011_PNAS,Sato2011}. Within arsenolite, As$_4$O$_6$, a reduction of $B$ from 7 to 4~GPa was observed upon helium insertion~\cite{Gunka_2015_CGaD,Sans_2016_PhysRevB}, however the results were questioned due to the lack of low-pressure experimental data~\cite{Sans_2016_PhysRevB}.  To study the way in which the mechanical properties of the A-site vacant perovskites considered herein change as they are stuffed with helium, we calculated their bulk moduli, shear moduli, and estimated their Vickers hardness using the Teter model at zero pressure. The results are provided in Table \ref{tab:bulk_moduli}. 

\begin{table}
\caption{Calculated bulk moduli (\textit{B}), shear moduli (\textit{G}) and Vickers hardness calculated with the Teter model  ($H_\text{v,Teter}$) of MX$_3$ and HeMX$_3$ phases optimized at 0 GPa. Experimental values are given in parentheses.}
    \begin{tabular}{ccccc}
    \hline
        Species & Space Group & $B$ [GPa] & $G$ [GPa] & $H_\text{v,Teter}$ [GPa] \\ 
    \hline \hline
        CaZrF$_6$ & $Fm\bar{3}m$ & 58.6 ($\sim$36~\cite{Hancock2015}) & 24.4 & 3.7  \\ 
        $\text{[He]}_2$[CaZr]F$_6$ & $Fm\bar{3}m$ & 61.6 ($\sim$47~\cite{Lloyd_2021_ChemofMat}) & 25.1 & 3.8  \\
        AlF$_3$  & $R\bar{3}c$ & 46.5 (39(8)~\cite{Stavrou_2015_JChemPhys}) & 66.8 & 10.1  \\ 
        HeAlF$_3$  & $R\bar{3}c$ & 131.1 & 82.7 & 12.5  \\ 
        GaF$_3$  & $R\bar{3}c$ & 56.3 (37(3)~\cite{Jorgensen_2010_HighPresRes}) & 49.2 & 7.4  \\ 
        HeGaF$_3$  & $R\bar{3}c$ & 109.0 & 60.4 & 9.1  \\ 
        InF$_3$  & $R\bar{3}c$ & 37.0 & 24.1 & 3.7  \\ 
        HeInF$_3$  & $R\bar{3}c$ & 38.4 & 29.2 & 4.4  \\ 
        ScF$_3$  & $Pm\bar{3}m$ & 95.8 (76(3)~\cite{Wei2020}) & 41.3 & 6.2  \\ 
        HeScF$_3$  & $P4/mbm$ & 96.1 & 43.6 & 6.5  \\ 
        ReO$_3$  & $Pm\bar{3}m$ & 237.9 (100(18)~\cite{Batlogg1984,Jorgensen2000}) & 125.1 & 19.0  \\ 
        HeReO$_3$ & $Pm\bar{3}m$ & 237.8 & 131.8 & 19.9  \\ 
        \hline 
	\label{tab:bulk_moduli}
    \end{tabular}\\
\end{table}

Comparison of our calculated bulk moduli for [CaZr]F$_6$ and [He]$_2$[CaZr]F$_6$ with experiments suggests that the computational protocol we employed underestimates their absolute values by 15-20~GPa, but is able to capture the increase in $B$ that occurs when the helium atoms fill the pores. Benchmarks using this same methodology showed that for a broad range of materials -- with $B$ as high as 434~GPa (for diamond) -- errors of 15~GPa were typical~\cite{Toher_2017_PhysRevMat}. Deviations of a few GPa in the measured values are not unusual between different experiments, but in some cases, differences of several tens or even 100~GPa were observed~\cite{Toher_2017_PhysRevMat}. 

The bulk moduli of HeMX$_3$ structures characterized by large cavities (larger than the vdW volume of helium), such as those containing indium, scandium and calcium-zirconium (Figure\ \ref{fig:phases}(f)), are predicted to be only slightly higher than their helium-free predecessors. However, perovskites that possessed smaller cavities, like GaF$_3$ and AlF$_3$, experienced a remarkable increase in their compressibility, with a large jump in $B$ as helium filled their pores. For example, our calculations suggest that the shear modulus of AlF$_3$ would increase from a value characteristic of tin (47~GPa) to a value characteristic of steel (131~GPa) upon helium insertion.  The (nearly equivalent) bulk moduli of ReO$_3$ and HeReO$_3$ quite high because the cubic $Pm\bar{3}m$ space group is naturally stiff, and they are also likely enhanced by the covalent character of the Re-O chemical bond~\cite{Gao2003, Xu2013}.  However, the bulk modulus of $Pm\bar{3}m$ ReO$_3$ was measured to be only 100(18) GPa~\cite{Batlogg1984, Jorgensen2000}, using a Birch-type equation of states, but assuming the pressure derivative $B_{0}^{'}$ to be zero, which may cause the large difference with the computed value. Moreover, the bulk modulus of ReO$_3$ is measured to drop to 43(1) GPa upon phase transition to $Im\bar{3}$~\cite{Jorgensen2004a}. Therefore, the presence of helium would prevent this phase transition, making HeReO$_3$ quite an incompressible material. Generally speaking the sheer moduli of these vacant A-site perovskites also increased upon helium insertion, but the increase was significantly smaller than for the bulk moduli, and all of the materials studied were quite soft.

\begin{figure}[H]
	\centering
		\includegraphics[width=3.5in]{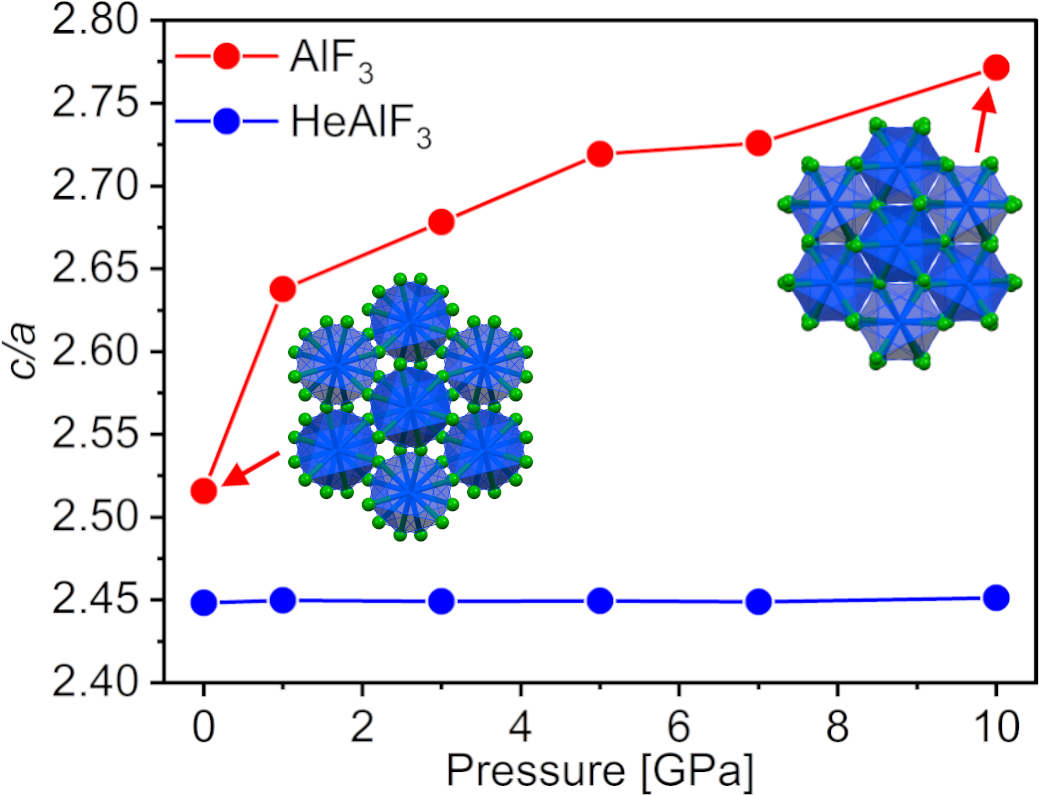}
	\caption{Response of the $c/a$ ratio of AlF$_3$ (red) and HeAlF$_3$ (blue) upon compression. The structural evolution of AlF$_3$ from 0 to 10~GPa is shown, illustrating how the structure changes from an open porous framework to a structure with a dense 8/3/c2 sphere packing.}
	\label{fig:ca_parameters}
\end{figure}

It is useful to consider the sphere-packing type of these perovskites, which can be characterized using three parameters, $k$/$m$/$fn$. These refer to the number of contacts per sphere, the mesh size (which is related to the number of edges or vertices in a shortest closed path of the sphere packing graph), and a symbol representing the crystal family and class, respectively~\cite{Fischer1973,Koch1995}. Typically, A-site vacant perovskites assume a rigid  sphere packing of the 8/3/c2 type~\cite{Sowa_1998_ActaCrysB} when compressed~\cite{Jorgensen_2010_HighPresRes}. This structural distortion is common for VF$_3$-type materials, and directly dependent on the nature of the cation. Evidence of the dense 8/3/c2 sphere packing can be obtained by monitoring the decrease of the $c/a$ ratio of the cell parameters as a consequence of the rotation of the octahedra upon compression, as illustrated for AlF$_3$ by the red circles in Figure\ \ref{fig:ca_parameters}. However, when helium enters the A-site vacancies, the dense 8/3/c2 sphere packing can be hampered, or even blocked, so that the $c/a$ ratio remains constant with increasing pressure (Figure\ \ref{fig:ca_parameters} -- blue circles). As a result the compressibility along the $a$ and $c$ axis becomes isotropic once helium is absorbed, a crystallographic feature that could be used as a diagnostic to characterize the insertion of helium in VF$_3$-type perovskites with X-ray diffraction. However, it is likely that some care should be taken on the rate and amount of compression, because the dense 8/3/c2 sphere packing may seal the pores before helium has time to diffuse in the perovskite.

\subsubsection{Electronic Structure and Bonding}
Helium insertion also influences the electronic structure of a material, behaving as a hard-sphere and localizing the electron density associated with the atoms comprising the host framework. Computations have shown that the intake of helium increases the band gap of the alkali oxides~\cite{Gao_2019_PhysRevMAt}, arsenolite~\cite{Sans_2016_PhysRevB}, high pressure sodium \cite{Dong_2017_NatureChem,Gao_2020_NationalSciRev}, and of mixtures of helium with water, ammonia or methane~\cite{Bai_2019_CommChem, Gao_2020_NationalSciRev}. The band gaps we calculate, which are likely to be somewhat too small as they were obtained with the non-hybrid optB88-vdW functional, generally increase upon helium incorporation, especially under pressure (Table S3). There are two exceptions to this trend. The first is HeGaF$_3$, whose calculated band gap is slightly smaller at 1~GPa as compared to that of GaF$_3$. The  Ga-F-Ga angles, associated with the degree of rotation of the octahedra, increase from 138.1$^{\circ}$ in GaF$_3$ to 169.4$^{\circ}$ in HeGaF$_3$, and this is likely the reason behind this trend~\cite{Prasanna2017}. HeReO$_3$ and ReO$_3$ are both metallic throughout the whole pressure range considered in this study, in agreement with experiments confirming the conductivity of ReO$_3$~\cite{Fujimori1992}. 

The high ionization energy and low electron affinity of helium make it the least reactive of all of the elements that are known.\cite{Moore_1971, Hotop_1985_JPhysChemRefData}
 With the exception of exotic gas-phase cations like HHe$^+$\cite{Yousif_1989_JPhysA,Partridge_1999a_MolecPhys,Bovino_2011_AstronAstroPhys}, which is very strongly bound as it is isoelectronic to H$_2$ \cite{Grochala_2018_FoundChem}, no systems are known where helium forms a genuine chemical bond with another atom.  Though many compounds have been predicted via computations~\cite{Liu_2018_NatureComm,Bai_2019_CommChem,Gao_2019_PhysRevMAt,Hou_2021_PhysRevB} -- most of them studied due to their potential relevance towards high-pressure materials and planetary science~\cite{Liu_2020_PhysRevX,Gao_2020_NationalSciRev, Zhang_2018_PhysRevLett} -- they do not show any evidence of covalent or ionic bond formation with the unreactive helium atoms. Besides the stabilization afforded by vdW interactions, in some ionic compounds under high pressure helium incorporation could be explained via its role as a damper of the Coulomb repulsion terms between ions of the same type~\cite{Liu_2018_NatureComm}.

As noted above, dispersion is certainly an important interaction between helium and the A-site vacant perovskites. But, could there be more? To answer this question, we first computed the atomic charges of HeAlF$_3$ using different techniques to divide up the electron density into contributions from the constituent atoms, and also using codes that employed both plane-wave and atom-centered basis sets (Table S5). 
Though the various methods disagreed on the magnitude and even the sign of the charge on the helium atom, they all agreed on one thing: the amount is small. For example, the Bader charges are, among all presented charges, the closest to the ideal formal charges of the atoms, despite underestimating them to +2.56 on Al, -0.85 on F and -0.02 on He. In $\alpha$-AlF$_3$, the charge of Al does not change, while F is only 0.005 more negative than in HeAlF$_3$. Thus, the helium atoms within HeAlF$_3$ can effectively be viewed as being neutral, so that they cannot participate in ionic bonding with the atoms comprising the perovskite framework.  
 
\begin{figure}[h!]
		\includegraphics[width=0.7\figurewidth]{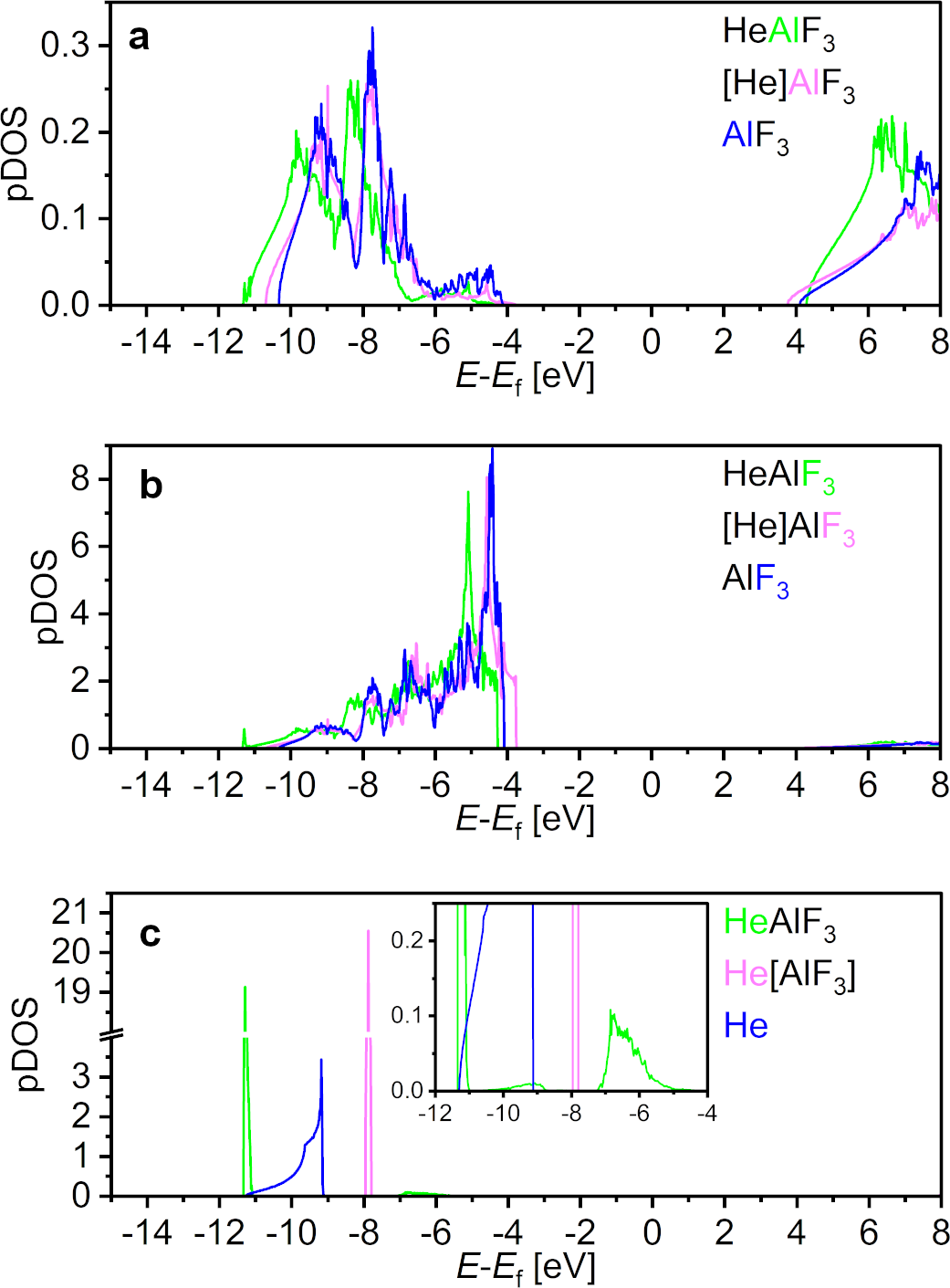}
	\caption{The atom projected densities of states (pDOS) of the (a) aluminium, (b) fluorine and (c) helium atoms in HeAlF$_3$, [He]AlF$_3$, $\alpha$-AlF$_3$ and hcp-He at 1~GPa. The zero of energy is set at the Fermi level (halfway between the top of the valence and bottom of the conduction bands).}
	\label{fig:Atomic_DOS}
\end{figure}

To search for potential signs of covalent bonding, we computed the atom projected density of states (pDOS) of HeAlF$_3$ at 1~GPa and compared it with the pDOS of two fictitious compounds that possessed the same geometry, but where either the helium atoms were removed, [He]AlF$_3$, or where the perovskite framework was deleted, He[AlF$_3$]. In addition, the pDOS of the 1~GPa $\alpha$-AlF$_3$ and hcp-He phases were obtained, and the results are plotted in Figure \ref{fig:Atomic_DOS}. The pDOS of both the Al and the F atoms in $\alpha$-AlF$_3$ and [He]AlF$_3$ are nearly the same, highlighting the structural similarity of the two lattices. Comparison of the pDOS associated with the framework atoms in HeAlF$_3$ and [He]AlF$_3$ shows that helium incorporation slightly shifts and broadens them, likely a result of the hard-sphere interactions that exert a chemical pressure on the Al and F atoms and opens up the band gap (Table S3). The DOS of helium reveals somewhat larger perturbations. The nearly isolated helium atoms within He[AlF$_3$] (separated by a distance of 3.63~\AA{}) are characterized by a $\delta$-function-like DOS around -8~eV, which is typical of a 0-dimensional system. As the atoms are pushed closer together to form the hcp-solid (He-He distances of 2.56~\AA{}), bands form and the DOS is broadened. Finally, the He PDOS within HeAlF$_3$ is also nearly $\delta$-function-like around -11~eV, except for a small contribution near -6~eV, which could also be explained by an overlap of the F and He atom-centered projection spheres. 

Furthermore, we calculated the COOP and COHP, which can diagnose if a certain interaction is bonding or anti-bonding~\cite{Hughbanks_1983_JACS}, determine the strength of that interaction~\cite{Dronskowski_1993_JPhysChem}, and uncover the orbitals that are its main constituents~\cite{Deringer_2011_JChemPhysA,Nelson_2020_JCompChem}.
Particularly useful is the negative of the integral of the COHP to the Fermi level (-ICOHP), which can be employed to quantify the covalent bond strength between two atoms. These chemical bonding descriptors were zero for both F-He and O-He atom pairs in all of the systems we considered at 1~GPa and 10~GPa. However, plots of the COOP and -COHP calculated for HeAlF$_3$ at 1~GPa (Figure S8(a)) hinted of a potential Al 3p and He 1s interaction. The -ICOHP obtained for this pair of atoms, separated by a distance of 3.144~\AA{}, suggested an interaction strength of $\sim$0.1~eV/bond (Figure S8(b)).
This value is similar to results obtained for unconventional C-H$\cdot\cdot\cdot$O hydrogen bonds that are weak and noncovalent in nature \cite{Deringer2014}.  
At 1~GPa the -ICOHPs between helium and the metal atoms in the remaining HeMX$_3$ compounds were negligible. At 10~GPa, however, the He-Ga -ICOHPs increased above 0.05~eV/bond as the distance between the atoms decreased (Figure S8(c)). These values resemble previous results obtained for He-H, He-N and He-O atom pairs at 400-500~GPa, which were attributed to an erroneous projection, expected with this technique for the short interatomic distances that occur at high pressures~\cite{Bai_2019_CommChem}. Given the small radii of the metal cations, the large distances between helium and the metal atom, and the inertness of this noble gas, we conclude that the small He-Al and He-Ga -ICOHPs calculated at 10~GPa are not indicative of a covalent interaction between these atoms, though other weak forces cannot be ruled out.

\section{Conclusions}
The recent pressure-induced insertion of helium into the vacant A-site perovskite [CaZr]F$_6$ has attracted much interest and tantalized with the prospect that other members of this perovskite family could encapsulate this noble gas as well~\cite{Hester_2017_JACS,Lloyd_2021_ChemofMat}. Because these experimental studies were unable to fully characterize the crystal structure of the [He]$_2$[CaZr]F$_6$ phase above $\sim$1~GPa, we employed evolutionary searches coupled with density functional theory (DFT) calculations to predict the most stable geometry at this pressure. The  $Fm\bar{3}m\rightarrow P2_1/c$ structural phase transition that occurs when [He]$_2$[CaZr]F$_6$ is squeezed was found to be driven by a decrease in the pressure-volume term to the enthalpy, which favored helium incorporation due to the attractive van der Waals interactions. A comparison of the sizes of the pores in the A-site vacant perovskite with the radius of helium showed that they are sufficiently large to accommodate the noble gas atoms. However, above 3~GPa the pores within [He]$_2$[CaZr]F$_6$ collapsed, coinciding with the amorphization of this phase that was observed in experiments. Though our DFT calculations overestimated the bulk moduli of the vacant and He-stuffed perovskite, they were able to capture how the bulk moduli increases upon helium insertion.

Systematic DFT calculations investigating the potential for helium incorporation into other A-site vacant perovskites, AlF$_3$, GaF$_3$, InF$_3$, ScF$_3$ and ReO$_3$ were performed. A topological analysis of the pores found within them revealed that some were spherical, whereas others were figure eight shaped, and their volume was compared with that of the helium atom at the same pressure. Evolutionary structure searches predicted the most stable helium-filled phases up to 10~GPa. Though all of the discovered structures were dynamically stable, only HeInF$_3$, HeScF$_3$ and HeReO$_3$ were predicted to be thermodynamically stable within specific pressure ranges based on the static-lattice enthalpy alone. Entropic contributions to the Gibbs free energy were found to favor the incorporation of helium into the perovskite at pressures where the volume this atom occupies within the pores was larger than within the highly compressible elemental phase. Thus, our DFT calculations showed that by tuning pressure and temperature conditions could be found that favored helium insertion into these A-site vacant perovskites. Helium incorporation into compounds with small pores, such as AlF$_3$ and GaF$_3$, increased their bulk moduli dramatically, and it hampered a structural distortion commonly observed in VF$_3$ type materials that is associated with rotation of their octahedra under compression. 

We hope our DFT results may guide the development of new synthesis pathways for the insertion of helium into A-site vacant perovskite materials, and will help elucidate their resulting mechanical and structural properties. 
Perovskite phases are also a major constituents of the Earth's mantle and the mantles of other rocky planets~\cite{Hirose_2017_Science} and the results presented here suggest that helium may occupy vacancies or defects found within them.~\cite{Verma2009, Gruninger2019} Thus, the quantity of helium that is stored within planetary interiors may be larger than previously believed~\cite{Stuart_2003_Nature, Jackson2017_Nature}.

\begin{acknowledgement}

Partial funding for this research is provided by the Center for Matter at Atomic Pressures (CMAP), a National Science Foundation (NSF) Physics Frontiers Center, under Award PHY-2020249. Calculations were performed at the Center for Computational Research at SUNY Buffalo \cite{ccr}. M.M. acknowledges the support of the NSF (DMR 1848141 and OAC-2117956), ACS PRF (50249-UNI6), and the California State University Research, Scholarship and Creative Activity (RSCA) award. We acknowledge Angn\`es Dewaele for useful discussions.

\end{acknowledgement}

\begin{suppinfo}

POTCARs used in this work, enthalpies of formation of [He]$_2$[CaZr]F$_6$ calculated with the PBE and optB88-vdW functionals,  band-gaps, enthalpies of formation of various phases as a function of pressure, Gibbs free energy changes as a function of temperature at fixed pressures, atomic charges of HeAlF$_3$,  electronic densities of states, phonon band structures, COOP and COHP results. The optimized structures are collected as \emph{cif} files in a separate folder.

\end{suppinfo}

\bibliography{racioppi-et-al_References.bib}

\providecommand{\latin}[1]{#1}
\makeatletter
\providecommand{\doi}
  {\begingroup\let\do\@makeother\dospecials
  \catcode`\{=1 \catcode`\}=2 \doi@aux}
\providecommand{\doi@aux}[1]{\endgroup\texttt{#1}}
\makeatother
\providecommand*\mcitethebibliography{\thebibliography}
\csname @ifundefined\endcsname{endmcitethebibliography}
  {\let\endmcitethebibliography\endthebibliography}{}
\begin{mcitethebibliography}{126}
\providecommand*\natexlab[1]{#1}
\providecommand*\mciteSetBstSublistMode[1]{}
\providecommand*\mciteSetBstMaxWidthForm[2]{}
\providecommand*\mciteBstWouldAddEndPuncttrue
  {\def\EndOfBibitem{\unskip.}}
\providecommand*\mciteBstWouldAddEndPunctfalse
  {\let\EndOfBibitem\relax}
\providecommand*\mciteSetBstMidEndSepPunct[3]{}
\providecommand*\mciteSetBstSublistLabelBeginEnd[3]{}
\providecommand*\EndOfBibitem{}
\mciteSetBstSublistMode{f}
\mciteSetBstMaxWidthForm{subitem}{(\alph{mcitesubitemcount})}
\mciteSetBstSublistLabelBeginEnd
  {\mcitemaxwidthsubitemform\space}
  {\relax}
  {\relax}

\bibitem[Jackson \latin{et~al.}(2010)Jackson, Carlson, Kurz, Kempton, Francis,
  and Blusztajn]{Jackson2010_Nature}
Jackson,~M.~G.; Carlson,~R.~W.; Kurz,~M.~D.; Kempton,~P.~D.; Francis,~D.;
  Blusztajn,~J. {Evidence for the Survival of the Oldest Terrestrial Mantle
  Reservoir}. \emph{Nature} \textbf{2010}, \emph{466}, 853--856\relax
\mciteBstWouldAddEndPuncttrue
\mciteSetBstMidEndSepPunct{\mcitedefaultmidpunct}
{\mcitedefaultendpunct}{\mcitedefaultseppunct}\relax
\EndOfBibitem
\bibitem[Hubbard and Militzer(2016)Hubbard, and
  Militzer]{Hubbard_2016_TheAstrophysJourn}
Hubbard,~W.~B.; Militzer,~B. {A Preliminary Jupiter Model}. \emph{Astrophys.
  J.} \textbf{2016}, \emph{820}, 13\relax
\mciteBstWouldAddEndPuncttrue
\mciteSetBstMidEndSepPunct{\mcitedefaultmidpunct}
{\mcitedefaultendpunct}{\mcitedefaultseppunct}\relax
\EndOfBibitem
\bibitem[Moore(1971)]{Moore_1971}
Moore,~C.~E. \emph{Atomic Energy Levels as Derived from the Analyses of Optical
  Spectra}; 1971; Vol.~1; p 358\relax
\mciteBstWouldAddEndPuncttrue
\mciteSetBstMidEndSepPunct{\mcitedefaultmidpunct}
{\mcitedefaultendpunct}{\mcitedefaultseppunct}\relax
\EndOfBibitem
\bibitem[Hotop and Lineberger(1985)Hotop, and
  Lineberger]{Hotop_1985_JPhysChemRefData}
Hotop,~H.; Lineberger,~W.~C. {Binding Energies in Atomic Negative Ions: II}.
  \emph{Journal of Physical and Chemical Reference Data} \textbf{1985},
  \emph{14}, 731--750\relax
\mciteBstWouldAddEndPuncttrue
\mciteSetBstMidEndSepPunct{\mcitedefaultmidpunct}
{\mcitedefaultendpunct}{\mcitedefaultseppunct}\relax
\EndOfBibitem
\bibitem[Hotokka \latin{et~al.}(1984)Hotokka, Kindstedt, Pyykk{\"{o}}, and
  Roos]{Hotokka_1984_MolecularPhysics}
Hotokka,~M.; Kindstedt,~T.; Pyykk{\"{o}},~P.; Roos,~B.~O. {On Bonding in
  Transition-Metal Helide Ions}. \emph{Mol. Phys.} \textbf{1984}, \emph{52},
  23--32\relax
\mciteBstWouldAddEndPuncttrue
\mciteSetBstMidEndSepPunct{\mcitedefaultmidpunct}
{\mcitedefaultendpunct}{\mcitedefaultseppunct}\relax
\EndOfBibitem
\bibitem[Koch \latin{et~al.}(1987)Koch, Frenking, Gauss, Cremer, and
  Collins]{Koch_1987_JACS}
Koch,~W.; Frenking,~G.; Gauss,~J.; Cremer,~D.; Collins,~J.~R. {Helium
  Chemistry: Theoretical Predictions and Experimental Challenge}. \emph{J. Am.
  Chem. Soc.} \textbf{1987}, \emph{109}, 5917--5934\relax
\mciteBstWouldAddEndPuncttrue
\mciteSetBstMidEndSepPunct{\mcitedefaultmidpunct}
{\mcitedefaultendpunct}{\mcitedefaultseppunct}\relax
\EndOfBibitem
\bibitem[Frenking \latin{et~al.}(1989)Frenking, Koch, Cremer, Gauss, and
  Liebman]{Frenking_1989_JPhysChem}
Frenking,~G.; Koch,~W.; Cremer,~D.; Gauss,~J.; Liebman,~J.~F. {Helium Bonding
  in Singly and Doubly Charged First-row Diatomic Cations HeX$_n$$^+$ (X=
  Li-Ne; n= 1, 2)}. \emph{J. Phys. Chem.} \textbf{1989}, \emph{93},
  3397--3410\relax
\mciteBstWouldAddEndPuncttrue
\mciteSetBstMidEndSepPunct{\mcitedefaultmidpunct}
{\mcitedefaultendpunct}{\mcitedefaultseppunct}\relax
\EndOfBibitem
\bibitem[Wilson \latin{et~al.}(2002)Wilson, Marsden, and {Von
  Nagy-Felsobuki}]{Wilson2002}
Wilson,~D.~J.; Marsden,~C.~J.; {Von Nagy-Felsobuki},~E.~I. {Ab Initio
  Structures and Stabilities of Doubly Charged Diatomic Metal Helides for the
  First Row Transition metals}. \emph{J. Phys. Chem. A} \textbf{2002},
  \emph{106}, 7348--7354\relax
\mciteBstWouldAddEndPuncttrue
\mciteSetBstMidEndSepPunct{\mcitedefaultmidpunct}
{\mcitedefaultendpunct}{\mcitedefaultseppunct}\relax
\EndOfBibitem
\bibitem[Li \latin{et~al.}(2005)Li, Mou, Chen, and Hu]{Li2005}
Li,~T.-H.; Mou,~C.-H.; Chen,~H.-R.; Hu,~W.-P. {Theoretical Prediction of Noble
  Gas Containing Anions FNgO$^-$ (Ng = He, Ar and Kr)}. \emph{J. Am. Chem.
  Soc.} \textbf{2005}, \emph{127}, 9241--9245\relax
\mciteBstWouldAddEndPuncttrue
\mciteSetBstMidEndSepPunct{\mcitedefaultmidpunct}
{\mcitedefaultendpunct}{\mcitedefaultseppunct}\relax
\EndOfBibitem
\bibitem[Hogness and Lunn(1925)Hogness, and Lunn]{Hogness1925}
Hogness,~T.~R.; Lunn,~E.~G. The Ionization of Hydrogen by Electron Impact as
  Interpreted by Positive Ray Analysis. \emph{Phys. Rev.} \textbf{1925},
  \emph{26}, 44--55\relax
\mciteBstWouldAddEndPuncttrue
\mciteSetBstMidEndSepPunct{\mcitedefaultmidpunct}
{\mcitedefaultendpunct}{\mcitedefaultseppunct}\relax
\EndOfBibitem
\bibitem[Saunders \latin{et~al.}(1994)Saunders, Jim{\'{e}}nez-V{\'{a}}zquez,
  {James Cross}, Mroczkowski, Gross, Giblin, and Poreda]{Saunders1994}
Saunders,~M.; Jim{\'{e}}nez-V{\'{a}}zquez,~H.~A.; {James Cross},~R.;
  Mroczkowski,~S.; Gross,~M.; Giblin,~D.~E.; Poreda,~R.~J. {Incorporation of
  Helium, Neon, Argon, Krypton, and Xenon into Fullerenes Using High Pressure}.
  \emph{J. Am. Chem. Soc.} \textbf{1994}, \emph{116}, 2193--2194\relax
\mciteBstWouldAddEndPuncttrue
\mciteSetBstMidEndSepPunct{\mcitedefaultmidpunct}
{\mcitedefaultendpunct}{\mcitedefaultseppunct}\relax
\EndOfBibitem
\bibitem[Saunders \latin{et~al.}(1996)Saunders, Cross,
  Jim{\'{e}}nez-V{\'{a}}zquez, Shimshi, and Khong]{Saunders1996}
Saunders,~M.; Cross,~R.~J.; Jim{\'{e}}nez-V{\'{a}}zquez,~H.~A.; Shimshi,~R.;
  Khong,~A. {Noble Gas Atoms Inside Fullerenes}. \emph{Science} \textbf{1996},
  \emph{271}, 1693--1697\relax
\mciteBstWouldAddEndPuncttrue
\mciteSetBstMidEndSepPunct{\mcitedefaultmidpunct}
{\mcitedefaultendpunct}{\mcitedefaultseppunct}\relax
\EndOfBibitem
\bibitem[Cappelletti \latin{et~al.}(2015)Cappelletti, Bartocci, Grandinetti,
  Falcinelli, Belpassi, Tarantelli, and Pirani]{Cappelletti2015}
Cappelletti,~D.; Bartocci,~A.; Grandinetti,~F.; Falcinelli,~S.; Belpassi,~L.;
  Tarantelli,~F.; Pirani,~F. {Experimental Evidence of Chemical Components in
  the Bonding of Helium and Neon with Neutral Molecules}. \emph{Chem. Eur. J.}
  \textbf{2015}, \emph{21}, 6234--6240\relax
\mciteBstWouldAddEndPuncttrue
\mciteSetBstMidEndSepPunct{\mcitedefaultmidpunct}
{\mcitedefaultendpunct}{\mcitedefaultseppunct}\relax
\EndOfBibitem
\bibitem[Frenking \latin{et~al.}(1988)Frenking, Koch, Gauss, and
  Cremer]{Frenking:1988a}
Frenking,~G.; Koch,~W.; Gauss,~J.; Cremer,~D. Stabilities and Nature of the
  Attractive Interactions in HeBeO, NeBeO, and ArBeO and a Comparison with
  Analogues NgLiF, NgBN, and NgLiH (Ng= He, Ar). A Theoretical Investigation.
  \emph{J. Am. Chem. Soc.} \textbf{1988}, \emph{110}, 8007--8016\relax
\mciteBstWouldAddEndPuncttrue
\mciteSetBstMidEndSepPunct{\mcitedefaultmidpunct}
{\mcitedefaultendpunct}{\mcitedefaultseppunct}\relax
\EndOfBibitem
\bibitem[Borocci \latin{et~al.}(2006)Borocci, Bronzolino, and
  Grandinetti]{Borocci2006}
Borocci,~S.; Bronzolino,~N.; Grandinetti,~F. {Neutral Helium Compounds:
  Theoretical Evidence for a Large Class of Polynuclear Complexes}. \emph{Chem.
  Eur. J.} \textbf{2006}, \emph{12}, 5033--5042\relax
\mciteBstWouldAddEndPuncttrue
\mciteSetBstMidEndSepPunct{\mcitedefaultmidpunct}
{\mcitedefaultendpunct}{\mcitedefaultseppunct}\relax
\EndOfBibitem
\bibitem[Borocci \latin{et~al.}(2020)Borocci, Grandinetti, Sanna, Antoniotti,
  and Nunzi]{Borocci2020}
Borocci,~S.; Grandinetti,~F.; Sanna,~N.; Antoniotti,~P.; Nunzi,~F. {Complexes
  of Helium with Neutral Molecules: Progress Toward a Quantitative Scale of
  Bonding Character}. \emph{J. Comput. Chem.} \textbf{2020}, \emph{41},
  1000--1011\relax
\mciteBstWouldAddEndPuncttrue
\mciteSetBstMidEndSepPunct{\mcitedefaultmidpunct}
{\mcitedefaultendpunct}{\mcitedefaultseppunct}\relax
\EndOfBibitem
\bibitem[Grochala(2012)]{Grochala:2012a}
Grochala,~W. A Metastable He-O Bond Inside a Ferroelectric Molecular cavity:
  (HeO)(LiF)$_2$. \emph{Phys. Chem. Chem. Phys.} \textbf{2012}, \emph{14},
  14860--14868\relax
\mciteBstWouldAddEndPuncttrue
\mciteSetBstMidEndSepPunct{\mcitedefaultmidpunct}
{\mcitedefaultendpunct}{\mcitedefaultseppunct}\relax
\EndOfBibitem
\bibitem[Wong(2000)]{Wong:2000a}
Wong,~M.~W. Prediction of a Metastable Helium Compound: HHeF. \emph{J. Am.
  Chem. Soc.} \textbf{2000}, \emph{122}, 6289--6290\relax
\mciteBstWouldAddEndPuncttrue
\mciteSetBstMidEndSepPunct{\mcitedefaultmidpunct}
{\mcitedefaultendpunct}{\mcitedefaultseppunct}\relax
\EndOfBibitem
\bibitem[Saha \latin{et~al.}(2019)Saha, Jana, Pan, Merino, and
  Chattaraj]{Saha2019}
Saha,~R.; Jana,~G.; Pan,~S.; Merino,~G.; Chattaraj,~P.~K. {How Far Can One Push
  the Noble Gases Towards Bonding?: A Personal Account}. \emph{Molecules}
  \textbf{2019}, \emph{24}, 1--23\relax
\mciteBstWouldAddEndPuncttrue
\mciteSetBstMidEndSepPunct{\mcitedefaultmidpunct}
{\mcitedefaultendpunct}{\mcitedefaultseppunct}\relax
\EndOfBibitem
\bibitem[Morinaka \latin{et~al.}(2013)Morinaka, Sato, Wakamiya, Nikawa,
  Mizorogi, Tanabe, Murata, Komatsu, Furukawa, Kato, Nagase, Akasaka, and
  Murata]{Morinaka2013}
Morinaka,~Y.; Sato,~S.; Wakamiya,~A.; Nikawa,~H.; Mizorogi,~N.; Tanabe,~F.;
  Murata,~M.; Komatsu,~K.; Furukawa,~K.; Kato,~T.; Nagase,~S.; Akasaka,~T.;
  Murata,~Y. {X-ray Observation of a Helium Atom and Placing a Nitrogen Atom
  Inside He@C60 and He@C70}. \emph{Nat. Commun.} \textbf{2013}, \emph{4},
  1--5\relax
\mciteBstWouldAddEndPuncttrue
\mciteSetBstMidEndSepPunct{\mcitedefaultmidpunct}
{\mcitedefaultendpunct}{\mcitedefaultseppunct}\relax
\EndOfBibitem
\bibitem[Grochala \latin{et~al.}(2007)Grochala, Hoffmann, Feng, and
  Ashcroft]{Grochala2007}
Grochala,~W.; Hoffmann,~R.; Feng,~J.; Ashcroft,~N.~W. {The Chemical Imagination
  at Work in Very Tight Places}. \emph{Angew. Chem., Int. Ed.} \textbf{2007},
  \emph{46}, 3620--3642\relax
\mciteBstWouldAddEndPuncttrue
\mciteSetBstMidEndSepPunct{\mcitedefaultmidpunct}
{\mcitedefaultendpunct}{\mcitedefaultseppunct}\relax
\EndOfBibitem
\bibitem[Miao \latin{et~al.}(2020)Miao, Sun, Zurek, and Lin]{Miao2020Nature}
Miao,~M.; Sun,~Y.; Zurek,~E.; Lin,~H. {Chemistry Under High Pressure}.
  \emph{Nat. Rev. Chem.} \textbf{2020}, \emph{4}, 508--527\relax
\mciteBstWouldAddEndPuncttrue
\mciteSetBstMidEndSepPunct{\mcitedefaultmidpunct}
{\mcitedefaultendpunct}{\mcitedefaultseppunct}\relax
\EndOfBibitem
\bibitem[Hilleke and Zurek(2023)Hilleke, and Zurek]{Hilleke2023}
Hilleke,~K.~P.; Zurek,~E. In \emph{Comprehensive Inorganic Chemistry III};
  Poeppelmeier,~K., Reedijk,~J., Eds.; Elsevier, 2023\relax
\mciteBstWouldAddEndPuncttrue
\mciteSetBstMidEndSepPunct{\mcitedefaultmidpunct}
{\mcitedefaultendpunct}{\mcitedefaultseppunct}\relax
\EndOfBibitem
\bibitem[Miao(2017)]{Miao2017Nature}
Miao,~M. {Helium Chemistry: React with Nobility}. \emph{Nat. Chem.}
  \textbf{2017}, \emph{9}, 409--410\relax
\mciteBstWouldAddEndPuncttrue
\mciteSetBstMidEndSepPunct{\mcitedefaultmidpunct}
{\mcitedefaultendpunct}{\mcitedefaultseppunct}\relax
\EndOfBibitem
\bibitem[Miao(2020)]{Miao_2020_FiC}
Miao,~M. {Noble Gases in Solid Compounds Show a Rich Display of Chemistry With
  Enough Pressure}. \emph{Front. Chem.} \textbf{2020}, \emph{8}, 1--8\relax
\mciteBstWouldAddEndPuncttrue
\mciteSetBstMidEndSepPunct{\mcitedefaultmidpunct}
{\mcitedefaultendpunct}{\mcitedefaultseppunct}\relax
\EndOfBibitem
\bibitem[Grochala(2018)]{Grochala_2018_FoundChem}
Grochala,~W. {On the Position of Helium and Neon in the Periodic Table of
  Elements}. \emph{Found. Chem.} \textbf{2018}, \emph{20}, 191--207\relax
\mciteBstWouldAddEndPuncttrue
\mciteSetBstMidEndSepPunct{\mcitedefaultmidpunct}
{\mcitedefaultendpunct}{\mcitedefaultseppunct}\relax
\EndOfBibitem
\bibitem[Zurek and Grochala(2015)Zurek, and Grochala]{Zurek2015}
Zurek,~E.; Grochala,~W. {Predicting Crystal Structures and Properties of Matter
  Under Extreme Conditions via Quantum Mechanics: The Pressure is On}.
  \emph{Phys. Chem. Chem. Phys.} \textbf{2015}, \emph{17}, 2917--2934\relax
\mciteBstWouldAddEndPuncttrue
\mciteSetBstMidEndSepPunct{\mcitedefaultmidpunct}
{\mcitedefaultendpunct}{\mcitedefaultseppunct}\relax
\EndOfBibitem
\bibitem[Zurek(2017)]{Zurek2017}
Zurek,~E. \emph{Handbook of Solid State Chemistry}; 2017; pp 571--605\relax
\mciteBstWouldAddEndPuncttrue
\mciteSetBstMidEndSepPunct{\mcitedefaultmidpunct}
{\mcitedefaultendpunct}{\mcitedefaultseppunct}\relax
\EndOfBibitem
\bibitem[Liu \latin{et~al.}(2015)Liu, Yao, and Klug]{Liu_2015_PhysRevB}
Liu,~H.; Yao,~Y.; Klug,~D.~D. {Stable Structures of He and H$_2$O at High
  Pressure}. \emph{Phys. Rev. B Condens. Matter Mater. Phys.} \textbf{2015},
  \emph{91}, 014102\relax
\mciteBstWouldAddEndPuncttrue
\mciteSetBstMidEndSepPunct{\mcitedefaultmidpunct}
{\mcitedefaultendpunct}{\mcitedefaultseppunct}\relax
\EndOfBibitem
\bibitem[Liu \latin{et~al.}(2020)Liu, Gao, Hermann, Wang, Miao, Pickard, Needs,
  Wang, Xing, and Sun]{Liu_2020_PhysRevX}
Liu,~C.; Gao,~H.; Hermann,~A.; Wang,~Y.; Miao,~M.; Pickard,~C.~J.;
  Needs,~R.~J.; Wang,~H.~T.; Xing,~D.; Sun,~J. {Plastic and Superionic Helium
  Ammonia Compounds under High Pressure and High Temperature}. \emph{Phys. Rev.
  X.} \textbf{2020}, \emph{10}, 21007\relax
\mciteBstWouldAddEndPuncttrue
\mciteSetBstMidEndSepPunct{\mcitedefaultmidpunct}
{\mcitedefaultendpunct}{\mcitedefaultseppunct}\relax
\EndOfBibitem
\bibitem[Shi \latin{et~al.}(2020)Shi, Cui, Hao, Xu, Wang, and
  Li]{Shi_2020_NatureComm}
Shi,~J.; Cui,~W.; Hao,~J.; Xu,~M.; Wang,~X.; Li,~Y. {Formation of
  Ammonia-Helium Compounds at High Pressure}. \emph{Nat. Commun.}
  \textbf{2020}, \emph{11}, 1--7\relax
\mciteBstWouldAddEndPuncttrue
\mciteSetBstMidEndSepPunct{\mcitedefaultmidpunct}
{\mcitedefaultendpunct}{\mcitedefaultseppunct}\relax
\EndOfBibitem
\bibitem[Liu \latin{et~al.}(2019)Liu, Gao, Wang, Needs, Pickard, Sun, Wang, and
  Xing]{Liu:2019a}
Liu,~C.; Gao,~H.; Wang,~Y.; Needs,~R.~J.; Pickard,~C.~J.; Sun,~J.; Wang,~H.~T.;
  Xing,~D. Multiple Superionic States in Helium-Water Compounds. \emph{Nature
  Phys.} \textbf{2019}, \emph{15}, 1065--1070\relax
\mciteBstWouldAddEndPuncttrue
\mciteSetBstMidEndSepPunct{\mcitedefaultmidpunct}
{\mcitedefaultendpunct}{\mcitedefaultseppunct}\relax
\EndOfBibitem
\bibitem[Gao \latin{et~al.}(2020)Gao, Liu, Hermann, Needs, Pickard, Wang, Xing,
  and Sun]{Gao_2020_NationalSciRev}
Gao,~H.; Liu,~C.; Hermann,~A.; Needs,~R.~J.; Pickard,~C.~J.; Wang,~H.~T.;
  Xing,~D.; Sun,~J. {Coexistence of Plastic and Partially Diffusive Phases in a
  Helium-methane Compound}. \emph{Natl. Sci. Rev.} \textbf{2020}, \emph{7},
  1540--1547\relax
\mciteBstWouldAddEndPuncttrue
\mciteSetBstMidEndSepPunct{\mcitedefaultmidpunct}
{\mcitedefaultendpunct}{\mcitedefaultseppunct}\relax
\EndOfBibitem
\bibitem[Monserrat \latin{et~al.}(2018)Monserrat, Martinez-Canales, Needs, and
  Pickard]{Monserrat_2018_PhysRevLett}
Monserrat,~B.; Martinez-Canales,~M.; Needs,~R.~J.; Pickard,~C.~J. {Helium-Iron
  Compounds at Terapascal Pressures}. \emph{Phys. Rev. Lett.} \textbf{2018},
  \emph{121}, 15301\relax
\mciteBstWouldAddEndPuncttrue
\mciteSetBstMidEndSepPunct{\mcitedefaultmidpunct}
{\mcitedefaultendpunct}{\mcitedefaultseppunct}\relax
\EndOfBibitem
\bibitem[Li \latin{et~al.}(2018)Li, Feng, Liu, Hao, Redfern, Lei, Liu, and
  Ma]{Li_2018_NatureComm}
Li,~Y.; Feng,~X.; Liu,~H.; Hao,~J.; Redfern,~S.~A.; Lei,~W.; Liu,~D.; Ma,~Y.
  {Route to High-energy Density Polymeric Nitrogen t-N via He-N Compounds}.
  \emph{Nat. Commun.} \textbf{2018}, \emph{9}, 1--7\relax
\mciteBstWouldAddEndPuncttrue
\mciteSetBstMidEndSepPunct{\mcitedefaultmidpunct}
{\mcitedefaultendpunct}{\mcitedefaultseppunct}\relax
\EndOfBibitem
\bibitem[Hou \latin{et~al.}(2021)Hou, Weng, Oganov, Shao, Gao, Dong, Wang,
  Tian, and Zhou]{Hou_2021_PhysRevB}
Hou,~J.; Weng,~X.~J.; Oganov,~A.~R.; Shao,~X.; Gao,~G.; Dong,~X.; Wang,~H.~T.;
  Tian,~Y.; Zhou,~X.~F. {Helium-nitrogen Mixtures at High Pressure}.
  \emph{Phys. Rev. B} \textbf{2021}, \emph{103}, L060102\relax
\mciteBstWouldAddEndPuncttrue
\mciteSetBstMidEndSepPunct{\mcitedefaultmidpunct}
{\mcitedefaultendpunct}{\mcitedefaultseppunct}\relax
\EndOfBibitem
\bibitem[Wei \latin{et~al.}(2019)Wei, Zhao, Zhang, Yan, Wei, and
  Peng]{Wei_2019_JAlloy}
Wei,~Q.; Zhao,~C.; Zhang,~M.; Yan,~H.; Wei,~B.; Peng,~X. {New Stable Structures
  of HeN$_3$ Predicted Using First-principles Calculations}. \emph{J. Alloys
  Compd.} \textbf{2019}, \emph{800}, 505--511\relax
\mciteBstWouldAddEndPuncttrue
\mciteSetBstMidEndSepPunct{\mcitedefaultmidpunct}
{\mcitedefaultendpunct}{\mcitedefaultseppunct}\relax
\EndOfBibitem
\bibitem[Dong \latin{et~al.}(2017)Dong, Oganov, Goncharov, Stavrou, Lobanov,
  Saleh, Qian, Zhu, Gatti, Deringer, Dronskowski, Zhou, Prakapenka,
  Kon{\^{o}}pkov{\'{a}}, Popov, Boldyrev, and Wang]{Dong_2017_NatureChem}
Dong,~X. \latin{et~al.}  {A Stable Compound of Helium and Sodium at High
  Pressure}. \emph{Nat. Chem.} \textbf{2017}, \emph{9}, 440--445\relax
\mciteBstWouldAddEndPuncttrue
\mciteSetBstMidEndSepPunct{\mcitedefaultmidpunct}
{\mcitedefaultendpunct}{\mcitedefaultseppunct}\relax
\EndOfBibitem
\bibitem[Liu \latin{et~al.}(2018)Liu, Botana, Hermann, Valdez, Zurek, Yan, Lin,
  and Miao]{Liu_2018_NatureComm}
Liu,~Z.; Botana,~J.; Hermann,~A.; Valdez,~S.; Zurek,~E.; Yan,~D.; Lin,~H.~Q.;
  Miao,~M.~S. {Reactivity of He with Ionic Compounds Under High Pressure}.
  \emph{Nat. Commun.} \textbf{2018}, \emph{9}, 1--10\relax
\mciteBstWouldAddEndPuncttrue
\mciteSetBstMidEndSepPunct{\mcitedefaultmidpunct}
{\mcitedefaultendpunct}{\mcitedefaultseppunct}\relax
\EndOfBibitem
\bibitem[Zhang \latin{et~al.}(2018)Zhang, Lv, Li, Feng, Lu, Redfern, Liu, Chen,
  and Ma]{Zhang_2018_PhysRevLett}
Zhang,~J.; Lv,~J.; Li,~H.; Feng,~X.; Lu,~C.; Redfern,~S.~A.; Liu,~H.; Chen,~C.;
  Ma,~Y. {Rare Helium-Bearing Compound FeO$_2$He Stabilized at Deep-Earth
  Conditions}. \emph{Phys. Rev. Lett.} \textbf{2018}, \emph{121}, 255703\relax
\mciteBstWouldAddEndPuncttrue
\mciteSetBstMidEndSepPunct{\mcitedefaultmidpunct}
{\mcitedefaultendpunct}{\mcitedefaultseppunct}\relax
\EndOfBibitem
\bibitem[Gao \latin{et~al.}(2019)Gao, Sun, Pickard, and
  Needs]{Gao_2019_PhysRevMAt}
Gao,~H.; Sun,~J.; Pickard,~C.~J.; Needs,~R.~J. {Prediction of Pressure-induced
  Stabilization of Noble-gas-atom Compounds with Alkali Oxides and Alkali
  Sulfides}. \emph{Phys. Rev. Mat.} \textbf{2019}, \emph{3}, 015002\relax
\mciteBstWouldAddEndPuncttrue
\mciteSetBstMidEndSepPunct{\mcitedefaultmidpunct}
{\mcitedefaultendpunct}{\mcitedefaultseppunct}\relax
\EndOfBibitem
\bibitem[Bai \latin{et~al.}(2019)Bai, Liu, Botana, Yan, Lin, Sun, Pickard,
  Needs, and Miao]{Bai_2019_CommChem}
Bai,~Y.; Liu,~Z.; Botana,~J.; Yan,~D.; Lin,~H.~Q.; Sun,~J.; Pickard,~C.~J.;
  Needs,~R.~J.; Miao,~M.~S. {Electrostatic Force Driven Helium Insertion Into
  Ammonia and Water Crystals Under Pressure}. \emph{Commun. Chem.}
  \textbf{2019}, \emph{2}, 1--7\relax
\mciteBstWouldAddEndPuncttrue
\mciteSetBstMidEndSepPunct{\mcitedefaultmidpunct}
{\mcitedefaultendpunct}{\mcitedefaultseppunct}\relax
\EndOfBibitem
\bibitem[Vos \latin{et~al.}(1992)Vos, Finger, Hemley, Hu, Mao, and
  Schouten]{Vos_1992_Nature}
Vos,~W.~L.; Finger,~L.~W.; Hemley,~R.~J.; Hu,~J.~Z.; Mao,~H.~K.;
  Schouten,~J.~A. {A High-pressure Van der Waals Compound in Solid
  Nitrogen-helium Mixtures}. \emph{Nature} \textbf{1992}, \emph{358},
  46--48\relax
\mciteBstWouldAddEndPuncttrue
\mciteSetBstMidEndSepPunct{\mcitedefaultmidpunct}
{\mcitedefaultendpunct}{\mcitedefaultseppunct}\relax
\EndOfBibitem
\bibitem[Zhang \latin{et~al.}(2017)Zhang, Cai, Bi, Zarifi, Terpstra, Zhang,
  Valy~Verdeny, Zurek, and Deemyad]{Zurek:2017f}
Zhang,~R.; Cai,~W.; Bi,~T.; Zarifi,~N.; Terpstra,~T.; Zhang,~C.;
  Valy~Verdeny,~Z.; Zurek,~E.; Deemyad,~S. Effects of Non-Hydrostatic Stress on
  Structural and Optoelectronic Properties of Methylammonium Lead Bromide
  Perovskite. \emph{J. Phys. Chem. Lett.} \textbf{2017}, \emph{8},
  3457--3465\relax
\mciteBstWouldAddEndPuncttrue
\mciteSetBstMidEndSepPunct{\mcitedefaultmidpunct}
{\mcitedefaultendpunct}{\mcitedefaultseppunct}\relax
\EndOfBibitem
\bibitem[Klotz \latin{et~al.}(2009)Klotz, Chervin, Munsch, and {Le
  Marchand}]{Klotz2009}
Klotz,~S.; Chervin,~J.~C.; Munsch,~P.; {Le Marchand},~G. {Hydrostatic limits of
  11 pressure transmitting media}. \emph{J. Phys. D Appl. Phys.} \textbf{2009},
  \emph{42}, 075413\relax
\mciteBstWouldAddEndPuncttrue
\mciteSetBstMidEndSepPunct{\mcitedefaultmidpunct}
{\mcitedefaultendpunct}{\mcitedefaultseppunct}\relax
\EndOfBibitem
\bibitem[Londono \latin{et~al.}(1988)Londono, Kuhs, and
  Finney]{Londono_1988_Nature}
Londono,~D.; Kuhs,~W.~F.; Finney,~J.~L. {Enclathration of Helium in Ice II: The
  First Helium Hydrate}. \emph{Nature} \textbf{1988}, \emph{332},
  141--142\relax
\mciteBstWouldAddEndPuncttrue
\mciteSetBstMidEndSepPunct{\mcitedefaultmidpunct}
{\mcitedefaultendpunct}{\mcitedefaultseppunct}\relax
\EndOfBibitem
\bibitem[Londono \latin{et~al.}(1992)Londono, Finney, and
  Kuhs]{Londono_1992_JChemPhys}
Londono,~D.; Finney,~J.~L.; Kuhs,~W.~F. {Formation, Stabilty, and Structure of
  Helium Hydrate at High Pressure}. \emph{J. Chem. Phys.} \textbf{1992},
  \emph{97}, 547--552\relax
\mciteBstWouldAddEndPuncttrue
\mciteSetBstMidEndSepPunct{\mcitedefaultmidpunct}
{\mcitedefaultendpunct}{\mcitedefaultseppunct}\relax
\EndOfBibitem
\bibitem[Yagi \latin{et~al.}(2007)Yagi, Iida, Hirai, Miyajima, Kikegawa, and
  Bunno]{Yagi_2007_PhysRevB}
Yagi,~T.; Iida,~E.; Hirai,~H.; Miyajima,~N.; Kikegawa,~T.; Bunno,~M.
  {High-pressure Behavior of a SiO$_2$ Clathrate Observed by Using Various
  Pressure Media}. \emph{Phys. Rev. B Condens. Matter Mater. Phys.}
  \textbf{2007}, \emph{75}, 174115\relax
\mciteBstWouldAddEndPuncttrue
\mciteSetBstMidEndSepPunct{\mcitedefaultmidpunct}
{\mcitedefaultendpunct}{\mcitedefaultseppunct}\relax
\EndOfBibitem
\bibitem[Shen \latin{et~al.}(2011)Shen, Mei, Prakapenka, Lazor, Sinogeikin,
  Meng, and Park]{Shen_2011_PNAS}
Shen,~G.; Mei,~Q.; Prakapenka,~V.~B.; Lazor,~P.; Sinogeikin,~S.; Meng,~Y.;
  Park,~C. {Effect of Helium on Structure and Compression Behavior of SiO$_2$
  Glass}. \emph{Proc. Natl. Acad. Sci. U.S.A.} \textbf{2011}, \emph{108},
  6004--6007\relax
\mciteBstWouldAddEndPuncttrue
\mciteSetBstMidEndSepPunct{\mcitedefaultmidpunct}
{\mcitedefaultendpunct}{\mcitedefaultseppunct}\relax
\EndOfBibitem
\bibitem[Sato \latin{et~al.}(2011)Sato, Funamori, and Yagi]{Sato2011}
Sato,~T.; Funamori,~N.; Yagi,~T. {Helium Penetrates Into Silica Glass and
  Reduces its Compressibility}. \emph{Nat. Commun.} \textbf{2011}, \emph{2},
  2--6\relax
\mciteBstWouldAddEndPuncttrue
\mciteSetBstMidEndSepPunct{\mcitedefaultmidpunct}
{\mcitedefaultendpunct}{\mcitedefaultseppunct}\relax
\EndOfBibitem
\bibitem[Gu{\'{n}}ka \latin{et~al.}(2015)Gu{\'{n}}ka, Dziubek, G{\l}adysiak,
  Dranka, Piechota, Hanfland, Katrusiak, and Zachara]{Gunka_2015_CGaD}
Gu{\'{n}}ka,~P.~A.; Dziubek,~K.~F.; G{\l}adysiak,~A.; Dranka,~M.; Piechota,~J.;
  Hanfland,~M.; Katrusiak,~A.; Zachara,~J. {Compressed Arsenolite As$_4$O$_6$
  and Its Helium Clathrate As$_4$O$_6${\textperiodcentered}2He}. \emph{Cryst.
  Growth Des.} \textbf{2015}, \emph{15}, 3740--3745\relax
\mciteBstWouldAddEndPuncttrue
\mciteSetBstMidEndSepPunct{\mcitedefaultmidpunct}
{\mcitedefaultendpunct}{\mcitedefaultseppunct}\relax
\EndOfBibitem
\bibitem[Sans \latin{et~al.}(2016)Sans, Manj{\'{o}}n, Popescu, Cuenca-Gotor,
  Gomis, Mu{\~{n}}oz, Rodr{\'{i}}guez-Hern{\'{a}}ndez, Contreras-Garc{\'{i}}a,
  Pellicer-Porres, Pereira, Santamar{\'{i}}a-P{\'{e}}rez, and
  Segura]{Sans_2016_PhysRevB}
Sans,~J.~A.; Manj{\'{o}}n,~F.~J.; Popescu,~C.; Cuenca-Gotor,~V.~P.; Gomis,~O.;
  Mu{\~{n}}oz,~A.; Rodr{\'{i}}guez-Hern{\'{a}}ndez,~P.;
  Contreras-Garc{\'{i}}a,~J.; Pellicer-Porres,~J.; Pereira,~A.~L.;
  Santamar{\'{i}}a-P{\'{e}}rez,~D.; Segura,~A. {Ordered Helium Trapping and
  Bonding in Compressed Arsenolite: Synthesis of
  As$_4$O$_6${\textperiodcentered}2He}. \emph{Phys. Rev. B} \textbf{2016},
  \emph{93}, 054102\relax
\mciteBstWouldAddEndPuncttrue
\mciteSetBstMidEndSepPunct{\mcitedefaultmidpunct}
{\mcitedefaultendpunct}{\mcitedefaultseppunct}\relax
\EndOfBibitem
\bibitem[Hester \latin{et~al.}(2017)Hester, {Dos Santos}, Molaison, Hancock,
  and Wilkinson]{Hester_2017_JACS}
Hester,~B.~R.; {Dos Santos},~A.~M.; Molaison,~J.~J.; Hancock,~J.~C.;
  Wilkinson,~A.~P. {Synthesis of Defect Perovskites
  (He$_{2-x}$$\square$$_x$)(CaZr)F6 by Inserting Helium into the Negative
  Thermal Expansion Material CaZrF$_6$}. \emph{J. Am. Chem. Soc.}
  \textbf{2017}, \emph{139}, 13284--13287\relax
\mciteBstWouldAddEndPuncttrue
\mciteSetBstMidEndSepPunct{\mcitedefaultmidpunct}
{\mcitedefaultendpunct}{\mcitedefaultseppunct}\relax
\EndOfBibitem
\bibitem[Lloyd \latin{et~al.}(2021)Lloyd, Hester, Baxter, Ma, Prakapenka,
  Tkachev, Park, and Wilkinson]{Lloyd_2021_ChemofMat}
Lloyd,~A.~J.; Hester,~B.~R.; Baxter,~S.~J.; Ma,~S.; Prakapenka,~V.~B.;
  Tkachev,~S.~N.; Park,~C.; Wilkinson,~A.~P. {Hybrid Double Perovskite
  Containing Helium: [He$_2$][CaZr]F$_6$}. \emph{Chem. Mater.} \textbf{2021},
  \emph{33}, 3132--3138\relax
\mciteBstWouldAddEndPuncttrue
\mciteSetBstMidEndSepPunct{\mcitedefaultmidpunct}
{\mcitedefaultendpunct}{\mcitedefaultseppunct}\relax
\EndOfBibitem
\bibitem[Lonie and Zurek(2011)Lonie, and Zurek]{Lonie2011}
Lonie,~D.~C.; Zurek,~E. {XtalOpt: An Open-source Evolutionary Algorithm for
  Crystal Structure Prediction}. \emph{Comput. Phys. Commun.} \textbf{2011},
  \emph{182}, 372--387\relax
\mciteBstWouldAddEndPuncttrue
\mciteSetBstMidEndSepPunct{\mcitedefaultmidpunct}
{\mcitedefaultendpunct}{\mcitedefaultseppunct}\relax
\EndOfBibitem
\bibitem[Falls \latin{et~al.}(2020)Falls, Avery, Wang, Hilleke, and
  Zurek]{Falls_2020_JPhysChemC}
Falls,~Z.; Avery,~P.; Wang,~X.; Hilleke,~K.~P.; Zurek,~E. {The XtalOpt
  Evolutionary Algorithm for Crystal Structure Prediction}. \emph{J. Phys.
  Chem. C} \textbf{2020}, \emph{125}, 1601--1620\relax
\mciteBstWouldAddEndPuncttrue
\mciteSetBstMidEndSepPunct{\mcitedefaultmidpunct}
{\mcitedefaultendpunct}{\mcitedefaultseppunct}\relax
\EndOfBibitem
\bibitem[Avery \latin{et~al.}(2019)Avery, Toher, Curtarolo, and
  Zurek]{Zurek:2018j}
Avery,~P.; Toher,~C.; Curtarolo,~S.; Zurek,~E. XtalOpt version r12: An
  Open-Source Evolutionary Algorithm for Crystal Structure Prediction.
  \emph{Comput. Phys. Commun.} \textbf{2019}, \emph{237}, 274--275\relax
\mciteBstWouldAddEndPuncttrue
\mciteSetBstMidEndSepPunct{\mcitedefaultmidpunct}
{\mcitedefaultendpunct}{\mcitedefaultseppunct}\relax
\EndOfBibitem
\bibitem[Avery and Zurek(2017)Avery, and Zurek]{Avery2017}
Avery,~P.; Zurek,~E. {RandSpg: An open-source Program for Generating Atomistic
  Crystal Structures with Specific Spacegroups}. \emph{Comput. Phys. Commun.}
  \textbf{2017}, \emph{213}, 208--216\relax
\mciteBstWouldAddEndPuncttrue
\mciteSetBstMidEndSepPunct{\mcitedefaultmidpunct}
{\mcitedefaultendpunct}{\mcitedefaultseppunct}\relax
\EndOfBibitem
\bibitem[Lonie and Zurek(2012)Lonie, and Zurek]{Lonie2012}
Lonie,~D.~C.; Zurek,~E. {Identifying Duplicate Crystal Structures: XtalComp, an
  Open-source Solution}. \emph{Comput. Phys. Commun.} \textbf{2012},
  \emph{183}, 690--697\relax
\mciteBstWouldAddEndPuncttrue
\mciteSetBstMidEndSepPunct{\mcitedefaultmidpunct}
{\mcitedefaultendpunct}{\mcitedefaultseppunct}\relax
\EndOfBibitem
\bibitem[Kresse and Hafner(1994)Kresse, and Hafner]{Kresse_1994_PhysRevB}
Kresse,~G.; Hafner,~J. {Ab Initio Molecular-dynamics Simulation of the
  Liquid-metalamorphous-semiconductor Transition in Germanium}. \emph{Phys.
  Rev. B} \textbf{1994}, \emph{49}, 14251--14269\relax
\mciteBstWouldAddEndPuncttrue
\mciteSetBstMidEndSepPunct{\mcitedefaultmidpunct}
{\mcitedefaultendpunct}{\mcitedefaultseppunct}\relax
\EndOfBibitem
\bibitem[Kresse and Joubert(1999)Kresse, and Joubert]{Kresse_1999_PhysRevB}
Kresse,~G.; Joubert,~D. {From Ultrasoft Pseudopotentials to the Projector
  Augmented-wave Method}. \emph{Phys. Rev. B Condens. Matter Mater. Phys.}
  \textbf{1999}, \emph{59}, 1758--1775\relax
\mciteBstWouldAddEndPuncttrue
\mciteSetBstMidEndSepPunct{\mcitedefaultmidpunct}
{\mcitedefaultendpunct}{\mcitedefaultseppunct}\relax
\EndOfBibitem
\bibitem[Perdew \latin{et~al.}(1996)Perdew, Burke, and Ernzerhof]{Perdew1996}
Perdew,~J.~P.; Burke,~K.; Ernzerhof,~M. {[ERRATA] Generalized Gradient
  Approximation Made Simple.} \emph{Phys. Rev. Lett.} \textbf{1996}, \emph{77},
  3865--3868\relax
\mciteBstWouldAddEndPuncttrue
\mciteSetBstMidEndSepPunct{\mcitedefaultmidpunct}
{\mcitedefaultendpunct}{\mcitedefaultseppunct}\relax
\EndOfBibitem
\bibitem[Perdew \latin{et~al.}(1996)Perdew, Burke, and Ernzerhof]{Perdew1996a}
Perdew,~J.~P.; Burke,~K.; Ernzerhof,~M. {Generalized Gradient Approximation
  Made Simple}. \emph{Phys. Rev. Lett.} \textbf{1996}, \emph{77},
  3865--3868\relax
\mciteBstWouldAddEndPuncttrue
\mciteSetBstMidEndSepPunct{\mcitedefaultmidpunct}
{\mcitedefaultendpunct}{\mcitedefaultseppunct}\relax
\EndOfBibitem
\bibitem[Dion \latin{et~al.}(2004)Dion, Rydberg, Schr{\"{o}}der, Langreth, and
  Lundqvist]{Dion_2004__PhysRevLett}
Dion,~M.; Rydberg,~H.; Schr{\"{o}}der,~E.; Langreth,~D.~C.; Lundqvist,~B.~I.
  {Van der Waals Density Functional for General Geometries}. \emph{Phys. Rev.
  Lett.} \textbf{2004}, \emph{92}, 246401\relax
\mciteBstWouldAddEndPuncttrue
\mciteSetBstMidEndSepPunct{\mcitedefaultmidpunct}
{\mcitedefaultendpunct}{\mcitedefaultseppunct}\relax
\EndOfBibitem
\bibitem[Klime{\v{s}} \latin{et~al.}(2010)Klime{\v{s}}, Bowler, and
  Michaelides]{Klimes_2010_JPhysCondMatt}
Klime{\v{s}},~J.; Bowler,~D.~R.; Michaelides,~A. {Chemical Accuracy for the Van
  der Waals Density Functional}. \emph{J. Phys. Condens. Matter.}
  \textbf{2010}, \emph{22}, 02201\relax
\mciteBstWouldAddEndPuncttrue
\mciteSetBstMidEndSepPunct{\mcitedefaultmidpunct}
{\mcitedefaultendpunct}{\mcitedefaultseppunct}\relax
\EndOfBibitem
\bibitem[Klime{\v{s}} \latin{et~al.}(2011)Klime{\v{s}}, Bowler, and
  Michaelides]{Klimes_2011_PhysRevB}
Klime{\v{s}},~J.; Bowler,~D.~R.; Michaelides,~A. {Van der Waals Density
  Functionals Applied to Solids}. \emph{Phys. Rev. B Condens. Matter Mater.
  Phys.} \textbf{2011}, \emph{83}, 195131\relax
\mciteBstWouldAddEndPuncttrue
\mciteSetBstMidEndSepPunct{\mcitedefaultmidpunct}
{\mcitedefaultendpunct}{\mcitedefaultseppunct}\relax
\EndOfBibitem
\bibitem[Bl{\"{o}}chl(1994)]{Blochl_1994_PhysRevB}
Bl{\"{o}}chl,~P.~E. {Projector Augmented-wave Method}. \emph{Phys. Rev. B}
  \textbf{1994}, \emph{50}, 17953--17979\relax
\mciteBstWouldAddEndPuncttrue
\mciteSetBstMidEndSepPunct{\mcitedefaultmidpunct}
{\mcitedefaultendpunct}{\mcitedefaultseppunct}\relax
\EndOfBibitem
\bibitem[Monkhorst and Pack(1976)Monkhorst, and Pack]{Monkhorst1977}
Monkhorst,~H.~J.; Pack,~J.~D. {Special points for Brillouin-zone integrations}.
  \emph{Phys. Rev. B} \textbf{1976}, \emph{13}, 5188--5192\relax
\mciteBstWouldAddEndPuncttrue
\mciteSetBstMidEndSepPunct{\mcitedefaultmidpunct}
{\mcitedefaultendpunct}{\mcitedefaultseppunct}\relax
\EndOfBibitem
\bibitem[Togo \latin{et~al.}(2008)Togo, Oba, and Tanaka]{Togo_2008_PhysRevB}
Togo,~A.; Oba,~F.; Tanaka,~I. {First-principles Calculations of the
  Ferroelastic Transition Between Rutile-type and CaCl$_2$-type SiO$_2$ at High
  Pressures}. \emph{Phys. Rev. B Condens. Matter Mater. Phys.} \textbf{2008},
  \emph{78}, 134106\relax
\mciteBstWouldAddEndPuncttrue
\mciteSetBstMidEndSepPunct{\mcitedefaultmidpunct}
{\mcitedefaultendpunct}{\mcitedefaultseppunct}\relax
\EndOfBibitem
\bibitem[Loubeyre \latin{et~al.}(1993)Loubeyre, Letoullec, Pinceaux, Mao, Hu,
  and Hemley]{Loubeyre_1993_PhysRevLettb}
Loubeyre,~P.; Letoullec,~R.; Pinceaux,~J.~P.; Mao,~H.~K.; Hu,~J.; Hemley,~R.~J.
  {Equation of State and Phase Diagram of Solid He from Single-crystal}.
  \emph{Phys. Rev. Lett.} \textbf{1993}, \emph{71}, 2272--2275\relax
\mciteBstWouldAddEndPuncttrue
\mciteSetBstMidEndSepPunct{\mcitedefaultmidpunct}
{\mcitedefaultendpunct}{\mcitedefaultseppunct}\relax
\EndOfBibitem
\bibitem[Clementi \latin{et~al.}(1967)Clementi, Raimondi, and
  Reinhardt]{Clementi1967}
Clementi,~E.; Raimondi,~D.~L.; Reinhardt,~W.~P. {Atomic screening constants
  from SCF functions. II. Atoms with 37 to 86 electrons}. \emph{J. Chem. Phys.}
  \textbf{1967}, \emph{47}, 1300--1307\relax
\mciteBstWouldAddEndPuncttrue
\mciteSetBstMidEndSepPunct{\mcitedefaultmidpunct}
{\mcitedefaultendpunct}{\mcitedefaultseppunct}\relax
\EndOfBibitem
\bibitem[MacRae \latin{et~al.}(2020)MacRae, Sovago, Cottrell, Galek, McCabe,
  Pidcock, Platings, Shields, Stevens, Towler, and Wood]{MacRae2020}
MacRae,~C.~F.; Sovago,~I.; Cottrell,~S.~J.; Galek,~P.~T.; McCabe,~P.;
  Pidcock,~E.; Platings,~M.; Shields,~G.~P.; Stevens,~J.~S.; Towler,~M.;
  Wood,~P.~A. {Mercury 4.0: From Visualization to Analysis, Design and
  Prediction}. \emph{J. Appl. Crystallogr.} \textbf{2020}, \emph{53},
  226--235\relax
\mciteBstWouldAddEndPuncttrue
\mciteSetBstMidEndSepPunct{\mcitedefaultmidpunct}
{\mcitedefaultendpunct}{\mcitedefaultseppunct}\relax
\EndOfBibitem
\bibitem[Rahm \latin{et~al.}(2016)Rahm, Hoffmann, and Ashcroft]{Rahm2016a}
Rahm,~M.; Hoffmann,~R.; Ashcroft,~N.~W. {Atomic and Ionic Radii of Elements
  1-96}. \emph{Chem. Eur. J.} \textbf{2016}, \emph{22}, 14625--14632\relax
\mciteBstWouldAddEndPuncttrue
\mciteSetBstMidEndSepPunct{\mcitedefaultmidpunct}
{\mcitedefaultendpunct}{\mcitedefaultseppunct}\relax
\EndOfBibitem
\bibitem[Alvarez(2013)]{Alvarez2013}
Alvarez,~S. {A Cartography of the Van der Waals Territories}. \emph{Dalton
  Trans.} \textbf{2013}, \emph{42}, 8617--8636\relax
\mciteBstWouldAddEndPuncttrue
\mciteSetBstMidEndSepPunct{\mcitedefaultmidpunct}
{\mcitedefaultendpunct}{\mcitedefaultseppunct}\relax
\EndOfBibitem
\bibitem[Bondi(1964)]{Bondi1964a}
Bondi,~A. {Van der Waals Volumes and Radii}. \emph{J. Phys. Chem.}
  \textbf{1964}, \emph{68}, 441--451\relax
\mciteBstWouldAddEndPuncttrue
\mciteSetBstMidEndSepPunct{\mcitedefaultmidpunct}
{\mcitedefaultendpunct}{\mcitedefaultseppunct}\relax
\EndOfBibitem
\bibitem[Rahm \latin{et~al.}(2020)Rahm, {\AA}ngqvist, Rahm, Erhart, and
  Cammi]{Rahm2020}
Rahm,~M.; {\AA}ngqvist,~M.; Rahm,~J.~M.; Erhart,~P.; Cammi,~R. {Non-Bonded
  Radii of the Atoms Under Compression}. \emph{ChemPhysChem} \textbf{2020},
  \emph{21}, 2441--2453\relax
\mciteBstWouldAddEndPuncttrue
\mciteSetBstMidEndSepPunct{\mcitedefaultmidpunct}
{\mcitedefaultendpunct}{\mcitedefaultseppunct}\relax
\EndOfBibitem
\bibitem[Cammi(2015)]{Cammi2015}
Cammi,~R. {A New Extension of the Polarizable Continuum Model: Toward a Quantum
  Chemical Description of Chemical Reactions at Extreme High Pressure}.
  \emph{J. Comput. Chem.} \textbf{2015}, \emph{36}, 2246--2259\relax
\mciteBstWouldAddEndPuncttrue
\mciteSetBstMidEndSepPunct{\mcitedefaultmidpunct}
{\mcitedefaultendpunct}{\mcitedefaultseppunct}\relax
\EndOfBibitem
\bibitem[sha()]{sharc}
\url{https://sharc.materialsmodeling.org/atoms_under_pressure/}, Accessed: Jan
  30, 2023\relax
\mciteBstWouldAddEndPuncttrue
\mciteSetBstMidEndSepPunct{\mcitedefaultmidpunct}
{\mcitedefaultendpunct}{\mcitedefaultseppunct}\relax
\EndOfBibitem
\bibitem[Hughbanks and Hoffmann(1983)Hughbanks, and
  Hoffmann]{Hughbanks_1983_JACS}
Hughbanks,~T.; Hoffmann,~R. {Chains of Trans-Edge-Sharing Molybdenum Octahedra:
  Metal-Metal Bonding in Extended Systems}. \emph{J. Am. Chem. Soc.}
  \textbf{1983}, \emph{105}, 3528--3537\relax
\mciteBstWouldAddEndPuncttrue
\mciteSetBstMidEndSepPunct{\mcitedefaultmidpunct}
{\mcitedefaultendpunct}{\mcitedefaultseppunct}\relax
\EndOfBibitem
\bibitem[Dronskowski and Blochl(1993)Dronskowski, and
  Blochl]{Dronskowski_1993_JPhysChem}
Dronskowski,~R.; Blochl,~P.~E. {Crystal Orbital Hamilton Population (COHP).
  Energy-Resolved Visualization of Chemical Bonding in Solid Based on
  Density-Functional Calculations}. \emph{J. Phys. Chem.} \textbf{1993},
  \emph{97}, 8617--8624\relax
\mciteBstWouldAddEndPuncttrue
\mciteSetBstMidEndSepPunct{\mcitedefaultmidpunct}
{\mcitedefaultendpunct}{\mcitedefaultseppunct}\relax
\EndOfBibitem
\bibitem[Deringer \latin{et~al.}(2011)Deringer, Tchougr{\'{e}}eff, and
  Dronskowski]{Deringer_2011_JChemPhysA}
Deringer,~V.~L.; Tchougr{\'{e}}eff,~A.~L.; Dronskowski,~R. {Crystal Orbital
  Hamilton Population (COHP) Analysis as Projected from Plane-wave Basis Sets}.
  \emph{J. Phys. Chem. A} \textbf{2011}, \emph{115}, 5461--5466\relax
\mciteBstWouldAddEndPuncttrue
\mciteSetBstMidEndSepPunct{\mcitedefaultmidpunct}
{\mcitedefaultendpunct}{\mcitedefaultseppunct}\relax
\EndOfBibitem
\bibitem[Nelson \latin{et~al.}(2020)Nelson, Ertural, George, Deringer, Hautier,
  and Dronskowski]{Nelson_2020_JCompChem}
Nelson,~R.; Ertural,~C.; George,~J.; Deringer,~V.~L.; Hautier,~G.;
  Dronskowski,~R. {LOBSTER: Local Orbital Projections, Atomic Charges, and
  Chemical-bonding Analysis from Projector-augmented-wave-based
  Density-functional Theory}. \emph{J. Comput. Chem.} \textbf{2020}, \emph{41},
  1931--1940\relax
\mciteBstWouldAddEndPuncttrue
\mciteSetBstMidEndSepPunct{\mcitedefaultmidpunct}
{\mcitedefaultendpunct}{\mcitedefaultseppunct}\relax
\EndOfBibitem
\bibitem[Tang \latin{et~al.}(2009)Tang, Sanville, and Henkelman]{Tang2009}
Tang,~W.; Sanville,~E.; Henkelman,~G. {A Grid-based Bader Analysis Algorithm
  Without Lattice Bias}. \emph{J. Phys. Condens. Matter.} \textbf{2009},
  \emph{21}, 084204\relax
\mciteBstWouldAddEndPuncttrue
\mciteSetBstMidEndSepPunct{\mcitedefaultmidpunct}
{\mcitedefaultendpunct}{\mcitedefaultseppunct}\relax
\EndOfBibitem
\bibitem[Philipsen \latin{et~al.}(2022)Philipsen, te~Velde, Baerends, Berger,
  de~Boeij, Franchini, Groeneveld, Kadantsev, Klooster, Kootstra, {M.C.W.M.
  Pols}, Raupach, Skachkov, Snijders, Verzijl, Gil, Thijssen, Wiesenekker,
  Peeples, Schreckenbach, and Ziegler]{Philipsen_2022_BAND}
Philipsen,~P. \latin{et~al.}  {BAND 2022.1, SCM, Theoretical Chemistry, Vrije
  Universiteit, Amsterdam, The Netherlands, http://www.scm.com}. 2022\relax
\mciteBstWouldAddEndPuncttrue
\mciteSetBstMidEndSepPunct{\mcitedefaultmidpunct}
{\mcitedefaultendpunct}{\mcitedefaultseppunct}\relax
\EndOfBibitem
\bibitem[{Fonseca Guerra} \latin{et~al.}(2004){Fonseca Guerra}, Handgraaf,
  Baerends, and Bickelhaupt]{FonsecaGuerra2004}
{Fonseca Guerra},~C.; Handgraaf,~J.~W.; Baerends,~E.~J.; Bickelhaupt,~F.~M.
  {Voronoi Deformation Density (VDD) Charges: Assessment of the Mulliken,
  Bader, Hirshfeld, Weinhold, and VDD Methods for Charge Analysis}. \emph{J.
  Comput. Chem.} \textbf{2004}, \emph{25}, 189--210\relax
\mciteBstWouldAddEndPuncttrue
\mciteSetBstMidEndSepPunct{\mcitedefaultmidpunct}
{\mcitedefaultendpunct}{\mcitedefaultseppunct}\relax
\EndOfBibitem
\bibitem[Perdew(1986)]{Perdew_1986_PhysRevB}
Perdew,~J.~P. {Density-functional Approximation for the Correlation Energy of
  the Inhomogeneous Electron Gas}. \emph{Phys. Rev. B} \textbf{1986},
  \emph{33}, 8822--8824\relax
\mciteBstWouldAddEndPuncttrue
\mciteSetBstMidEndSepPunct{\mcitedefaultmidpunct}
{\mcitedefaultendpunct}{\mcitedefaultseppunct}\relax
\EndOfBibitem
\bibitem[Becke(1988)]{Becke_1988_PhysRevA}
Becke,~A.~D. {Density-functional Exchange-energy Approximation with Correct
  Asymptotic Behavior}. \emph{Phys. Rev. A.} \textbf{1988}, \emph{38},
  3098--3100\relax
\mciteBstWouldAddEndPuncttrue
\mciteSetBstMidEndSepPunct{\mcitedefaultmidpunct}
{\mcitedefaultendpunct}{\mcitedefaultseppunct}\relax
\EndOfBibitem
\bibitem[Caldeweyher \latin{et~al.}(2020)Caldeweyher, Mewes, Ehlert, and
  Grimme]{Caldeweyher_2020_PCCP}
Caldeweyher,~E.; Mewes,~J.~M.; Ehlert,~S.; Grimme,~S. {Extension and Evaluation
  of the D4 London-dispersion Model for Periodic Systems}. \emph{Phys. Chem.
  Chem. Phys.} \textbf{2020}, \emph{22}, 8499--8512\relax
\mciteBstWouldAddEndPuncttrue
\mciteSetBstMidEndSepPunct{\mcitedefaultmidpunct}
{\mcitedefaultendpunct}{\mcitedefaultseppunct}\relax
\EndOfBibitem
\bibitem[Toher \latin{et~al.}(2017)Toher, Oses, Plata, Hicks, Rose, Levy, {De
  Jong}, Asta, Fornari, {Buongiorno Nardelli}, and
  Curtarolo]{Toher_2017_PhysRevMat}
Toher,~C.; Oses,~C.; Plata,~J.~J.; Hicks,~D.; Rose,~F.; Levy,~O.; {De
  Jong},~M.; Asta,~M.; Fornari,~M.; {Buongiorno Nardelli},~M.; Curtarolo,~S.
  {Combining the AFLOW GIBBS and Elastic Libraries to Efficiently and Robustly
  Screen Thermomechanical Properties of Solids}. \emph{Phys. Rev. Mat.}
  \textbf{2017}, \emph{1}, 015401\relax
\mciteBstWouldAddEndPuncttrue
\mciteSetBstMidEndSepPunct{\mcitedefaultmidpunct}
{\mcitedefaultendpunct}{\mcitedefaultseppunct}\relax
\EndOfBibitem
\bibitem[Teter(1998)]{Teter1998}
Teter,~D.~M. {Computational Alchemy: The Search for New Superhard Materials}.
  \emph{MRS Bulletin} \textbf{1998}, \emph{23}, 22--27\relax
\mciteBstWouldAddEndPuncttrue
\mciteSetBstMidEndSepPunct{\mcitedefaultmidpunct}
{\mcitedefaultendpunct}{\mcitedefaultseppunct}\relax
\EndOfBibitem
\bibitem[Avery \latin{et~al.}(2019)Avery, Wang, Oses, Gossett, Proserpio,
  Toher, Curtarolo, and Zurek]{Zurek:2019b}
Avery,~P.; Wang,~X.; Oses,~C.; Gossett,~E.; Proserpio,~D.~M.; Toher,~C.;
  Curtarolo,~S.; Zurek,~E. Predicting Superhard Materials via a Machine
  Learning Informed Evolutionary Structure Search. \emph{npj Comput. Mater.}
  \textbf{2019}, \emph{89}, 1--11\relax
\mciteBstWouldAddEndPuncttrue
\mciteSetBstMidEndSepPunct{\mcitedefaultmidpunct}
{\mcitedefaultendpunct}{\mcitedefaultseppunct}\relax
\EndOfBibitem
\bibitem[Hancock \latin{et~al.}(2015)Hancock, Chapman, Halder, Morelock,
  Kaplan, Gallington, Bongiorno, Han, Zhou, and Wilkinson]{Hancock2015}
Hancock,~J.~C.; Chapman,~K.~W.; Halder,~G.~J.; Morelock,~C.~R.; Kaplan,~B.~S.;
  Gallington,~L.~C.; Bongiorno,~A.; Han,~C.; Zhou,~S.; Wilkinson,~A.~P. {Large
  Negative Thermal Expansion and Anomalous Behavior on Compression in Cubic
  ReO$_3$-Type A$^{II}$B$^{IV}$F$_6$: CaZrF$_6$ and CaHfF$_6$}. \emph{Chem.
  Mater.} \textbf{2015}, \emph{27}, 3912--3918\relax
\mciteBstWouldAddEndPuncttrue
\mciteSetBstMidEndSepPunct{\mcitedefaultmidpunct}
{\mcitedefaultendpunct}{\mcitedefaultseppunct}\relax
\EndOfBibitem
\bibitem[Hoppe and Kissel(1984)Hoppe, and Kissel]{Hoppe_1984_JFluoChem}
Hoppe,~R.; Kissel,~D. {Zur Kenntnis von AlF$_3$ und InF$_3$}. \emph{J. Fluor.
  Chem.} \textbf{1984}, \emph{24}, 327--340\relax
\mciteBstWouldAddEndPuncttrue
\mciteSetBstMidEndSepPunct{\mcitedefaultmidpunct}
{\mcitedefaultendpunct}{\mcitedefaultseppunct}\relax
\EndOfBibitem
\bibitem[J{\o}rgensen \latin{et~al.}(2010)J{\o}rgensen, Olsen, and
  Gerward]{Jorgensen_2010_HighPresRes}
J{\o}rgensen,~J.~E.; Olsen,~J.~S.; Gerward,~L. {Compression Mechanism of
  GaF$_3$ and FeF$_3$: A High-pressure X-ray Diffraction Study}. \emph{High
  Press. Res.} \textbf{2010}, \emph{30}, 634--642\relax
\mciteBstWouldAddEndPuncttrue
\mciteSetBstMidEndSepPunct{\mcitedefaultmidpunct}
{\mcitedefaultendpunct}{\mcitedefaultseppunct}\relax
\EndOfBibitem
\bibitem[Stavrou \latin{et~al.}(2015)Stavrou, Zaug, Bastea, Crowhurst,
  Goncharov, Radousky, Armstrong, Roberts, and Plaue]{Stavrou_2015_JChemPhys}
Stavrou,~E.; Zaug,~J.~M.; Bastea,~S.; Crowhurst,~J.~C.; Goncharov,~A.~F.;
  Radousky,~H.~B.; Armstrong,~M.~R.; Roberts,~S.~K.; Plaue,~J.~W. {Equations of
  State of Anhydrous AlF$_3$ and AlI$_3$: Modeling of Extreme Condition Halide
  Chemistry}. \emph{J. Chem. Phys.} \textbf{2015}, \emph{142}, 214506\relax
\mciteBstWouldAddEndPuncttrue
\mciteSetBstMidEndSepPunct{\mcitedefaultmidpunct}
{\mcitedefaultendpunct}{\mcitedefaultseppunct}\relax
\EndOfBibitem
\bibitem[K{\"{o}}nig \latin{et~al.}(2010)K{\"{o}}nig, Scholz, Scheurell,
  Heidemann, Buchem, Unger, and Kemnitz]{Konig_2010_JFluoChem}
K{\"{o}}nig,~R.; Scholz,~G.; Scheurell,~K.; Heidemann,~D.; Buchem,~I.;
  Unger,~W.~E.; Kemnitz,~E. {Spectroscopic Characterization of Crystalline
  AlF$_3$ Phases}. \emph{J. Fluor. Chem.} \textbf{2010}, \emph{131},
  91--97\relax
\mciteBstWouldAddEndPuncttrue
\mciteSetBstMidEndSepPunct{\mcitedefaultmidpunct}
{\mcitedefaultendpunct}{\mcitedefaultseppunct}\relax
\EndOfBibitem
\bibitem[Krahl and Kemnitz(2017)Krahl, and Kemnitz]{Krahl_2017_CatSCiandTech}
Krahl,~T.; Kemnitz,~E. {Aluminium Fluoride-the Strongest Solid Lewis Acid:
  Structure and Reactivity}. \emph{Catal. Sci. Technol.} \textbf{2017},
  \emph{7}, 773--796\relax
\mciteBstWouldAddEndPuncttrue
\mciteSetBstMidEndSepPunct{\mcitedefaultmidpunct}
{\mcitedefaultendpunct}{\mcitedefaultseppunct}\relax
\EndOfBibitem
\bibitem[J{\o}rgensen \latin{et~al.}(2004)J{\o}rgensen, Marshall, Smith, Olsen,
  and Gerward]{Jorgensen2004a}
J{\o}rgensen,~J.~E.; Marshall,~W.~G.; Smith,~R.~I.; Olsen,~J.~S.; Gerward,~L.
  {High-pressure Neutron Powder Diffraction Study of the $Im\bar{3}$ Phase of
  ReO$_3$}. \emph{J. Appl. Crystallogr.} \textbf{2004}, \emph{37},
  857--861\relax
\mciteBstWouldAddEndPuncttrue
\mciteSetBstMidEndSepPunct{\mcitedefaultmidpunct}
{\mcitedefaultendpunct}{\mcitedefaultseppunct}\relax
\EndOfBibitem
\bibitem[Greve \latin{et~al.}(2010)Greve, Martin, Lee, Chupas, Chapman, and
  Wilkinson]{Greve_2010_JACS}
Greve,~B.~K.; Martin,~K.~L.; Lee,~P.~L.; Chupas,~P.~J.; Chapman,~K.~W.;
  Wilkinson,~A.~P. {Pronounced Negative Thermal Expansion from a Simple
  Structure: Cubic ScF$_3$}. \emph{J. Am. Chem. Soc.} \textbf{2010},
  \emph{132}, 15496--15498\relax
\mciteBstWouldAddEndPuncttrue
\mciteSetBstMidEndSepPunct{\mcitedefaultmidpunct}
{\mcitedefaultendpunct}{\mcitedefaultseppunct}\relax
\EndOfBibitem
\bibitem[Aleksandrov \latin{et~al.}(2002)Aleksandrov, Voronov, Vtyurin,
  Goryainov, Zamkova, Zinenko, and Krylov]{Aleksandrov_2002_ExpTheoPhys}
Aleksandrov,~K.~S.; Voronov,~V.~N.; Vtyurin,~A.~N.; Goryainov,~S.~V.;
  Zamkova,~N.~G.; Zinenko,~V.~I.; Krylov,~A.~S. {Lattice Dynamics and
  Hydrostatic-pressure-induced Phase Transitions in ScF$_3$}. \emph{J. Exp.
  Theor. Phys.} \textbf{2002}, \emph{94}, 977--984\relax
\mciteBstWouldAddEndPuncttrue
\mciteSetBstMidEndSepPunct{\mcitedefaultmidpunct}
{\mcitedefaultendpunct}{\mcitedefaultseppunct}\relax
\EndOfBibitem
\bibitem[J{\o}rgensen \latin{et~al.}(2000)J{\o}rgensen, {Staun Olsen}, and
  Gerward]{Jorgensen2000}
J{\o}rgensen,~J.~E.; {Staun Olsen},~J.; Gerward,~L. {Phase Transitions in
  ReO$_3$ Studied by High-pressure X-ray Diffraction}. \emph{J. Appl.
  Crystallogr.} \textbf{2000}, \emph{33}, 279--284\relax
\mciteBstWouldAddEndPuncttrue
\mciteSetBstMidEndSepPunct{\mcitedefaultmidpunct}
{\mcitedefaultendpunct}{\mcitedefaultseppunct}\relax
\EndOfBibitem
\bibitem[Glazer(1972)]{Glazer1972}
Glazer,~A.~M. {The Classification of Tilted Octahedra in Perovskites}.
  \emph{Acta Crystallogr. B Struct. Cryst.} \textbf{1972}, \emph{28},
  3384--3392\relax
\mciteBstWouldAddEndPuncttrue
\mciteSetBstMidEndSepPunct{\mcitedefaultmidpunct}
{\mcitedefaultendpunct}{\mcitedefaultseppunct}\relax
\EndOfBibitem
\bibitem[Stoumpos and Kanatzidis(2015)Stoumpos, and
  Kanatzidis]{Stoumpos_2015_AccChemRes}
Stoumpos,~C.~C.; Kanatzidis,~M.~G. {The Renaissance of Halide Perovskites and
  Their Evolution as Emerging Semiconductors}. \emph{Acc. Chem. Res.}
  \textbf{2015}, \emph{48}, 2791--2802\relax
\mciteBstWouldAddEndPuncttrue
\mciteSetBstMidEndSepPunct{\mcitedefaultmidpunct}
{\mcitedefaultendpunct}{\mcitedefaultseppunct}\relax
\EndOfBibitem
\bibitem[Akkerman and Manna(2020)Akkerman, and
  Manna]{Akkerman_2020_ACSEnergyLett}
Akkerman,~Q.~A.; Manna,~L. {What Defines a Halide Perovskite?} \emph{ACS Energy
  Lett.} \textbf{2020}, \emph{5}, 604--610\relax
\mciteBstWouldAddEndPuncttrue
\mciteSetBstMidEndSepPunct{\mcitedefaultmidpunct}
{\mcitedefaultendpunct}{\mcitedefaultseppunct}\relax
\EndOfBibitem
\bibitem[Wang \latin{et~al.}(2022)Wang, Gu{\'{e}}gan, and Frapper]{Wang2022a}
Wang,~B.; Gu{\'{e}}gan,~F.; Frapper,~G. {Putting Xenon and Nitrogen Under
  Pressure: Towards New Layered and Two-dimensional Nitrogen Allotropes with
  Crown Ether-like Nanopores}. \emph{J. Mater. Chem. C} \textbf{2022},
  \emph{10}, 10374--10381\relax
\mciteBstWouldAddEndPuncttrue
\mciteSetBstMidEndSepPunct{\mcitedefaultmidpunct}
{\mcitedefaultendpunct}{\mcitedefaultseppunct}\relax
\EndOfBibitem
\bibitem[Sun \latin{et~al.}(2016)Sun, Dacek, Ong, Hautier, Jain, Richards,
  Gamst, Persson, and Ceder]{Sun2016}
Sun,~W.; Dacek,~S.~T.; Ong,~S.~P.; Hautier,~G.; Jain,~A.; Richards,~W.~D.;
  Gamst,~A.~C.; Persson,~K.~A.; Ceder,~G. {The Thermodynamic Scale of Inorganic
  Crystalline Metastability}. \emph{Sci. Adv.} \textbf{2016}, \emph{2}\relax
\mciteBstWouldAddEndPuncttrue
\mciteSetBstMidEndSepPunct{\mcitedefaultmidpunct}
{\mcitedefaultendpunct}{\mcitedefaultseppunct}\relax
\EndOfBibitem
\bibitem[Wei \latin{et~al.}(2020)Wei, Tan, Cai, Phillips, {Da Silva}, Kibble,
  and Dove]{Wei2020}
Wei,~Z.; Tan,~L.; Cai,~G.; Phillips,~A.~E.; {Da Silva},~I.; Kibble,~M.~G.;
  Dove,~M.~T. {Colossal Pressure-Induced Softening in Scandium Fluoride}.
  \emph{Physical Review Letters} \textbf{2020}, \emph{124}, 255502\relax
\mciteBstWouldAddEndPuncttrue
\mciteSetBstMidEndSepPunct{\mcitedefaultmidpunct}
{\mcitedefaultendpunct}{\mcitedefaultseppunct}\relax
\EndOfBibitem
\bibitem[Batlogg \latin{et~al.}(1984)Batlogg, Maines, Greenblatt, and
  Digregorio]{Batlogg1984}
Batlogg,~B.; Maines,~R.~G.; Greenblatt,~M.; Digregorio,~S. {Novel p-V
  relationship in ReO3 under pressure}. \emph{Physical Review B} \textbf{1984},
  \emph{29}, 3762--3764\relax
\mciteBstWouldAddEndPuncttrue
\mciteSetBstMidEndSepPunct{\mcitedefaultmidpunct}
{\mcitedefaultendpunct}{\mcitedefaultseppunct}\relax
\EndOfBibitem
\bibitem[Gao \latin{et~al.}(2003)Gao, He, Wu, Liu, Yu, Li, Zhang, and
  Tian]{Gao2003}
Gao,~F.; He,~J.; Wu,~E.; Liu,~S.; Yu,~D.; Li,~D.; Zhang,~S.; Tian,~Y. {Hardness
  of Covalent Crystals}. \emph{Phys. Rev. Lett.} \textbf{2003}, \emph{91},
  015502\relax
\mciteBstWouldAddEndPuncttrue
\mciteSetBstMidEndSepPunct{\mcitedefaultmidpunct}
{\mcitedefaultendpunct}{\mcitedefaultseppunct}\relax
\EndOfBibitem
\bibitem[Xu \latin{et~al.}(2013)Xu, Wang, and Tian]{Xu2013}
Xu,~B.; Wang,~Q.; Tian,~Y. {Bulk Modulus for Polar Covalent Crystals}.
  \emph{Sci. Rep.} \textbf{2013}, \emph{3}, 1--7\relax
\mciteBstWouldAddEndPuncttrue
\mciteSetBstMidEndSepPunct{\mcitedefaultmidpunct}
{\mcitedefaultendpunct}{\mcitedefaultseppunct}\relax
\EndOfBibitem
\bibitem[Fischer(1973)]{Fischer1973}
Fischer,~W. {Existenzbedingungen Homogener Kugelpackungen zu Kubischen
  Gitterkomplexen mit Weniger als Drei Freiheitsgraden}. \emph{Z. fur Krist. -
  New Cryst. Struct.} \textbf{1973}, \emph{138}, 129--146\relax
\mciteBstWouldAddEndPuncttrue
\mciteSetBstMidEndSepPunct{\mcitedefaultmidpunct}
{\mcitedefaultendpunct}{\mcitedefaultseppunct}\relax
\EndOfBibitem
\bibitem[Koch and Fischer(1995)Koch, and Fischer]{Koch1995}
Koch,~E.; Fischer,~W. {Sphere Packings with Three Contacts per Sphere and the
  Problem of the Least Dense Sphere Packing}. \emph{Z. fur Krist. - New Cryst.
  Struct.} \textbf{1995}, \emph{210}, 407--414\relax
\mciteBstWouldAddEndPuncttrue
\mciteSetBstMidEndSepPunct{\mcitedefaultmidpunct}
{\mcitedefaultendpunct}{\mcitedefaultseppunct}\relax
\EndOfBibitem
\bibitem[Sowa and Ahsbahs(1998)Sowa, and Ahsbahs]{Sowa_1998_ActaCrysB}
Sowa,~H.; Ahsbahs,~H. {Pressure-Induced Octahedron Strain in VF$_3$-Type
  Compounds}. \emph{Acta. Crystallogr. B. Struct. Sci. Cryst. Eng. Mater.}
  \textbf{1998}, \emph{54}, 578--584\relax
\mciteBstWouldAddEndPuncttrue
\mciteSetBstMidEndSepPunct{\mcitedefaultmidpunct}
{\mcitedefaultendpunct}{\mcitedefaultseppunct}\relax
\EndOfBibitem
\bibitem[Prasanna \latin{et~al.}(2017)Prasanna, Gold-Parker, Leijtens, Conings,
  Babayigit, Boyen, Toney, and McGehee]{Prasanna2017}
Prasanna,~R.; Gold-Parker,~A.; Leijtens,~T.; Conings,~B.; Babayigit,~A.;
  Boyen,~H.~G.; Toney,~M.~F.; McGehee,~M.~D. {Band Gap Tuning via Lattice
  Contraction and Octahedral Tilting in Perovskite Materials for
  Photovoltaics}. \emph{J. Am. Chem. Soc.} \textbf{2017}, \emph{139},
  11117--11124\relax
\mciteBstWouldAddEndPuncttrue
\mciteSetBstMidEndSepPunct{\mcitedefaultmidpunct}
{\mcitedefaultendpunct}{\mcitedefaultseppunct}\relax
\EndOfBibitem
\bibitem[Fujimori(1992)]{Fujimori1992}
Fujimori,~A. {Electronic Structure of Metallic Oxides: Band-gap Closure and
  Valence Control}. \emph{J. Phys. Chem. Solids} \textbf{1992}, \emph{53},
  1595--1602\relax
\mciteBstWouldAddEndPuncttrue
\mciteSetBstMidEndSepPunct{\mcitedefaultmidpunct}
{\mcitedefaultendpunct}{\mcitedefaultseppunct}\relax
\EndOfBibitem
\bibitem[Yousif and Mitchell(1989)Yousif, and Mitchell]{Yousif_1989_JPhysA}
Yousif,~F.~B.; Mitchell,~J.~B. {Recombination and Excitation of HeH$^+$}.
  \emph{Phys. Rev. A.} \textbf{1989}, \emph{40}, 4318--4321\relax
\mciteBstWouldAddEndPuncttrue
\mciteSetBstMidEndSepPunct{\mcitedefaultmidpunct}
{\mcitedefaultendpunct}{\mcitedefaultseppunct}\relax
\EndOfBibitem
\bibitem[Partridge and Bauschlicher(1999)Partridge, and
  Bauschlicher]{Partridge_1999a_MolecPhys}
Partridge,~H.; Bauschlicher,~C.~W. {The Dissociation Energies of He$_2$, HeH,
  and ArH: A Bond Function Study}. \emph{Mol. Phys.} \textbf{1999}, \emph{96},
  705--710\relax
\mciteBstWouldAddEndPuncttrue
\mciteSetBstMidEndSepPunct{\mcitedefaultmidpunct}
{\mcitedefaultendpunct}{\mcitedefaultseppunct}\relax
\EndOfBibitem
\bibitem[Bovino \latin{et~al.}(2011)Bovino, Tacconi, Gianturco, and
  Galli]{Bovino_2011_AstronAstroPhys}
Bovino,~S.; Tacconi,~M.; Gianturco,~F.~A.; Galli,~D. {Ion Chemistry in the
  Early Universe: Revisiting the Role of HeH$^+$ with New Quantum
  Calculations}. \emph{Astron. Astrophys.} \textbf{2011}, \emph{529},
  1--5\relax
\mciteBstWouldAddEndPuncttrue
\mciteSetBstMidEndSepPunct{\mcitedefaultmidpunct}
{\mcitedefaultendpunct}{\mcitedefaultseppunct}\relax
\EndOfBibitem
\bibitem[Deringer \latin{et~al.}(2014)Deringer, Englert, and
  Dronskowski]{Deringer2014}
Deringer,~V.~L.; Englert,~U.; Dronskowski,~R. {Covalency of Hydrogen Bonds in
  Solids Revisited}. \emph{Chem. Comm.} \textbf{2014}, \emph{50},
  11547--11549\relax
\mciteBstWouldAddEndPuncttrue
\mciteSetBstMidEndSepPunct{\mcitedefaultmidpunct}
{\mcitedefaultendpunct}{\mcitedefaultseppunct}\relax
\EndOfBibitem
\bibitem[Hirose \latin{et~al.}(2017)Hirose, Sinmyo, and
  Hernlund]{Hirose_2017_Science}
Hirose,~K.; Sinmyo,~R.; Hernlund,~J. {Perovskite in Earth's Deep Interior}.
  \emph{Science} \textbf{2017}, \emph{358}, 734--738\relax
\mciteBstWouldAddEndPuncttrue
\mciteSetBstMidEndSepPunct{\mcitedefaultmidpunct}
{\mcitedefaultendpunct}{\mcitedefaultseppunct}\relax
\EndOfBibitem
\bibitem[Verma and Karki(2009)Verma, and Karki]{Verma2009}
Verma,~A.~K.; Karki,~B.~B. {Ab initio investigations of native and protonic
  point defects in Mg$_2$SiO$_4$ polymorphs under high pressure}. \emph{Earth
  and Planetary Science Letters} \textbf{2009}, \emph{285}, 140--149\relax
\mciteBstWouldAddEndPuncttrue
\mciteSetBstMidEndSepPunct{\mcitedefaultmidpunct}
{\mcitedefaultendpunct}{\mcitedefaultseppunct}\relax
\EndOfBibitem
\bibitem[Gr{\"{u}}ninger \latin{et~al.}(2019)Gr{\"{u}}ninger, Liu, Siegel,
  {Boffa Ballaran}, Katsura, Senker, and Frost]{Gruninger2019}
Gr{\"{u}}ninger,~H.; Liu,~Z.; Siegel,~R.; {Boffa Ballaran},~T.; Katsura,~T.;
  Senker,~J.; Frost,~D.~J. {Oxygen Vacancy Ordering in Aluminous Bridgmanite in
  the Earth's Lower Mantle}. \emph{Geophysical Research Letters} \textbf{2019},
  \emph{46}, 8731--8740\relax
\mciteBstWouldAddEndPuncttrue
\mciteSetBstMidEndSepPunct{\mcitedefaultmidpunct}
{\mcitedefaultendpunct}{\mcitedefaultseppunct}\relax
\EndOfBibitem
\bibitem[Stuart \latin{et~al.}(2003)Stuart, Lass-Evans, Fitton, and
  Ellam]{Stuart_2003_Nature}
Stuart,~F.~M.; Lass-Evans,~S.; Fitton,~J.~G.; Ellam,~R.~M. {High 3He/4He Ratios
  in Picritic Basalts from Baffin Island and the Role of a Mixed Reservoir in
  Mantle Plumes}. \emph{Nature} \textbf{2003}, \emph{424}, 57--59\relax
\mciteBstWouldAddEndPuncttrue
\mciteSetBstMidEndSepPunct{\mcitedefaultmidpunct}
{\mcitedefaultendpunct}{\mcitedefaultseppunct}\relax
\EndOfBibitem
\bibitem[Jackson \latin{et~al.}(2017)Jackson, Konter, and
  Becker]{Jackson2017_Nature}
Jackson,~M.~G.; Konter,~J.~G.; Becker,~T.~W. {Primordial Helium Entrained by
  the Hottest Mantle Plumes}. \emph{Nature} \textbf{2017}, \emph{542},
  340--343\relax
\mciteBstWouldAddEndPuncttrue
\mciteSetBstMidEndSepPunct{\mcitedefaultmidpunct}
{\mcitedefaultendpunct}{\mcitedefaultseppunct}\relax
\EndOfBibitem
\bibitem[ccr()]{ccr}
\emph{Center for Computational Research, University at Buffalo.}
  http://hdl.handle.net/10477/79221 (accessed 2023-02-09)\relax
\mciteBstWouldAddEndPuncttrue
\mciteSetBstMidEndSepPunct{\mcitedefaultmidpunct}
{\mcitedefaultendpunct}{\mcitedefaultseppunct}\relax
\EndOfBibitem
\end{mcitethebibliography}

\end{document}